\documentclass[%
superscriptaddress,
aps,
pra,
twocolumn
]{revtex4-2}

\usepackage{amssymb}
\usepackage[dvipdfmx]{graphicx}% Include figure files
\usepackage{sidecap}
\usepackage{dcolumn}% Align table columns on decimal point
\usepackage{bm}% bold math
\usepackage{nicefrac}
\usepackage{amsthm}
\usepackage{amsmath}
\usepackage{hyperref}
\usepackage{color}

\renewcommand*{\eqref}[1]{Eq.~(\ref{#1})}

\newcommand{\mbfalpha}{\bm{\alpha}}

\newcommand{\mbfx}{{\bf x}}%
\newcommand{\mbfy}{{\bf y}}
\newcommand{\mbfz}{\bm{z}}

\newcommand{\mbfa}{\bm{a}}
\newcommand{\mbfb}{\bm{b}}
\newcommand{\hp}{\hat{p}}
\newcommand{\hq}{\hat{q}}

\newcommand{\mbfp}{\bm{p}}
\newcommand{\mbfq}{\bm{q}}
\newcommand{\hmbfp}{\bm{\hat{p}}}
\newcommand{\hmbfq}{\bm{\hat{q}}}
\newcommand{\mbfd}{\bm{d}}
\newcommand{\mbfg}{\bm{g}}

\newcommand{\mbfn}{{\bf n}} 
\newcommand{\hpsi}{\hat{\psi}} \newcommand{\hpsid}{\hat{\psi}^\dagger}

\newcommand{\mbfR}{\bm{\mathrm{R}}}
\newcommand{\hmbfR}{\bm{\hat{R}}}

\newcommand{\mbfM}{{\bf M}}
\newcommand{\mbfJ}{\bm{\mathrm{J}}}

\newcommand{\mbfRq}{{\bf R}_\mathrm{q}}
\newcommand{\mbfRp}{{\bf R}_\mathrm{p}}
\newcommand{\mbfU}{\bm{\mathrm{U}}}
\newcommand{\mbfW}{\bm{\mathrm{W}}}

\newcommand{\ha}{\hat{a}}

\newcommand{\Tr}{\text{Tr}}

\graphicspath{{figure/}}
\begin{document}

%\preprint{APS/123-QED}

\title{
  Variational quantum algorithm for Gaussian discrete solitons and their boson sampling
}% Force line breaks with \\
%\thanks{Some notes}%

\author{Claudio Conti}%
 \email{claudio.conti@uniroma1.it}
\affiliation{%
Department of Physics, University Sapienza, P.le Aldo Moro 5, 00185 Rome, Italy
% Authors' institution and/or address\\
% This line break forced with \textbackslash\textbackslash
}%
\affiliation{%
  Institute for Complex Systems, National Research Council (ISC-CNR),
  Via dei Taurini 19, 00185 Rome, Italy
}%
\affiliation{%
Research Center Enrico Fermi (CREF), Via Panisperna 89a, 00184 Rome, Italy
% Authors' institution and/or address\\
% This line break forced with \textbackslash\textbackslash
}%

 \homepage{https://www.complexlight.org}

\date{\today}% It is always \today, today,
             %  but any date may be explicitly specified

\begin{abstract}
In the context of quantum information, highly nonlinear regimes, such as those supporting solitons, are marginally investigated. We miss general methods for quantum solitons,
although they can act as entanglement generators or as self-organized quantum processors.  We develop a computational approach that uses a neural network as a variational ansatz for quantum solitons in an array of waveguides. By training the resulting phase space quantum machine learning model, we find different soliton solutions varying the number of particles and interaction strength. We consider Gaussian states that enable measuring the degree of entanglement and sampling the probability distribution of many-particle events. We also determine the probability of generating particle pairs and unveil that soliton bound states emit correlated pairs. These results may have a role in boson sampling with nonlinear systems and in quantum processors for entangled nonlinear waves.
\end{abstract}

%\keywords{Suggested keywords}%Use showkeys class option if keyword
                              %display desired
\maketitle
A soliton is a non-perturbative solution of a classical nonlinear wave-equation; it may describe mean-field states of atoms (as in Bose-Einstein condensation) or photons (as in nonlinear optics)~\cite{KivsharBook}.
From a quantum mechanical perspective, a soliton may correspond to a coherent state; however, the nonlinearity may induce squeezing or non-Gaussianity~\cite{Conti2022}.
The quantum properties of solitons inspired experimental investigations,
as quantum non-demolition, squeezing~\cite{Friberg92,Drummond93, Silberhorn01,Yu:01} and photon bound states~\cite{Liang2018}.
Authors reported on theoretical studies on the soliton quantum features, as evaporation and breathing~\cite{Leuchs1999,Lee2005,Lai2009,villari2018,Malomed20,Martynov2020}.
Following these investigations, one can argue that solitons are new tools for fundamental many-body phenomena and sources of highly non-classical states.

However, solitons are overlooked in quantum technologies, as they require specialized theoretical methods~\cite{Potasek1987,LaiHauFirst1989, Wright91,Scott1994,Yao95} that do not appear compatible with modern quantum information. On the other hand, quantum optical protocols mainly employ linear circuits, and we ask if nonlinear waves may have a role in quantum computing~\cite{Marcucci2019b,Herrera2020,Silva2021,Pu2021,LopezPastor2021}, or --reciprocally-- if quantum processors can generate quantum solitons. An intriguing opportunity arises from the fact that linear circuits for the quantum advantage in boson sampling (BS)~\cite{Tillmann2012, Broome2013, Spring2013, Carolan2014, Spagnolo2014,Zhong2020} adopt the same devices used to observe discrete solitons~\cite{Lederer2008,Torner2011,KivsharBook}. BS in a linear optical circuit is ``intermediate'' between a many-body system not performing computation and a universal quantum computer~\cite{Aaronson2013}. To make a conceptual step, we can consider nonlinear phenomena in BS.
A soliton is a self-organized nonlinear channel.  Is the probability of specific events affected by these nonlinear channels? Can we do BS with solitons?

To merge solitons and quantum information, we need novel methodologies for quantum protocols on many-body states bounded by nonlinearity. Recently, neural networks (NNs) have been introduced to study many-body models and quantum circuits~\cite{Carleo2019}. Quantum machine learning (QML), i.e., the development and training of parameterized quantum circuits~\cite{Broughton2020, Marcucci:20,Marquardt2021,Tacchino2021,Kardashin2021}, is a technique to approximate high-dimensional ground states~\cite{Huang2021}. In QML, one considers quantum circuits depending on trainable parameters, such as rotation angles for qubits or degree of squeezing for continuous variables~\cite{SchuldPetruccioneBook}. Applications include quantum classifiers~\cite{Lloyd2014, daguanno2021}, or variational quantum algorithms~\cite{Vicentini2019, Cerezo2021} for optimization and quantum simulations.

Here we develop a quantum variational ansatz in the phase space for quantum soliton states.
The ansatz corresponds to a trainable BS quantum processor~\cite{Arrazola2021,Hoch2021} and can synthesize various soliton solutions. The trained NN is used to compute their quantum features, showing that soliton bound-states exhibit entanglement.
Also, the quantum variational algorithm is used to sample the probability distribution at specific patterns of output particles (i.e., BS) that reveal the generation of correlated bosons. The approach mixing phase space methods and machine learning turns out to be a remarkable tool for studying quantum solitons, and to predict and quantify novel effects not accessible with conventional techniques.
% These results may open the way to experiments with nonlinear waveguide arrays, and to the use of solitons as building blocks for quantum information.

This manuscript is organized as follows. In Sections I and II, we review the basics of phase space representations, characteristic functions, and outline the link with conventional NN models. In sections from III to IX, we show how a Gaussian state can be reinterpreted as a stack of linear layers with an Gaussian activation function; we discuss how to map different quantum operators to layers in a NN.
Then, in sections X and XI,  we represent the Gaussian ansatz for a quantum soliton as a quantum processor with trainable parameters and compute different soliton solutions. Finally, in sections XII and XIII, we use the trained neural network for various quantum features, as entanglement and BS. Conclusions are drawn in Section~\ref{sec:conclusion}.
\section{Quantum phase space and neural networks}
In the phase space, a state is represented by a function of several variables, which we encode in a real vector $\mbfx$.
For a $n$-body system, $\mbfx$ is a vector of $N=2\,n$ real variables.
Among the many representations~\cite{BarnettBook}, we consider the characteristic function $\chi(\mbfx)$.

For Gaussian states~\cite{X.2007}, in symmetric ordering,
\begin{equation}
  \label{eq:chiGaussian1}
  \chi(\mbfx)= \exp\left(-\frac{1}{4}\mbfx\cdot \mbfg \cdot \mbfx+\imath\, \mbfx\cdot \mbfd \right)\;,
\end{equation}
with $\mbfg$ the $N\times N$ covariance matrix, and $\mbfd$ the $N\times 1$ displacement vector. $\chi(\mbfx)$ is a complex function  $\chi(\mbfx)=\chi_R(\mbfx)+\imath \chi_I(\mbfx)$ with real part $\chi_{R}$ and imaginary part $\chi_{I}$.
Any many-body state is represented by a couple of real functions $\chi_{R,I}$ of $N$ real variables.
As NN models typically deal with real-valued quantities, in the following we will
represent the characteristic functions by models with two real outputs
corresponding to $\chi_R$ and $\chi_I$.
The resulting NN model includes linear transformations representing gates,
as displacements or interferometers, followed by a nonlinear activation, which corresponds to computing the characteristic functions (e.g, a Gaussian function). This maps the quantum state in a conventional NN architecture that can be trained by well-known algorithms.
 %------------------
\section{The characteristic function}\label{sec:PQMLchptCFRV}
We introduce the complex vector $\mbfz$ with components
\begin{equation}
  \mbfz=(z_0,z_0,\ldots,z_{n-1},z_0^*,z_1^*,\ldots,z_{n-1}^*)\;,
\end{equation}
which include the complex $z_j$ components and their conjugates $z_j^*$.
Given the density matrix $\rho$, the characteristic
function is~\cite{BarnettBook}
\begin{equation}
    \chi(\mbfz,\mathcal{P})
    =\Tr\left\{\rho e^{
    \sum_{k} \left( z_k\hat a^\dagger_k - z^*_k\hat a_k \right)}\right\}e^{\sum_k\frac{\mathcal{P}}{2}z_k z_k^*}\;,
        \label{eq:generalchi}
  \end{equation}
In Eq.~(\ref{eq:generalchi}) the variable $\mathcal{P}\in\{0,1,-1\}$, refers to the
adopted operator ordering.
The characteristic function depends on $\mbfz$ and the ordering index $\mathcal{P}$.
$\mathcal{P}$ is important when determining the
expected values as derivatives of $\chi$.

For an observable $\hat{O}$ we have
\begin{equation}
  \langle \hat{O}\rangle=\Tr\left(\rho \hat{O}\right).
\end{equation}

The expected value of a combination of the annihilation and creation operators is
\begin{equation}
    \langle {\left(\hat a_j^\dagger\right)}^m {\left(\hat a_k\right)}^n
\rangle_\mathcal{P}=
\left.{\left(\frac{\partial}{\partial z_j}\right)}^m
{\left(-\frac{\partial}{\partial z_k^*}\right)}^n
\chi(\mbfz,\mathcal{P})\right\vert_{\mbfz=0}\;,
\end{equation}
with $m$ and $n$ integers.

The mean value of the field operator
\begin{equation}
    \langle \hat a_k\rangle = -\frac{\partial \chi}{\partial z_k^*},
  \end{equation}
  does not depend on the ordering index $\mathcal{P}$, as also
  \begin{equation}
    \langle \hat a^\dagger_k\rangle = \frac{\partial \chi}{\partial z_k}.
  \end{equation}

The mean particle number of mode $k$, $\hat{n}_k=\ha_k^\dagger \ha_k$ is
\begin{equation}
  \langle \hat a_k^\dagger \hat a_k\rangle_\mathcal{P} =\left. -\frac{\partial^2 \chi(\mbfz,\mathcal{P})}
    {\partial z_k\partial z_k^*}\right\vert_{\mbfz=0}\;.
  \end{equation}
For the symmetric ordering $\mathcal{P}=0$, one has
  \begin{equation}
    \langle \hat a_k^\dagger \hat a_k\rangle_{\mathcal{P}=0}=
    \langle \hat a_k^\dagger \hat a_k+\hat a_k \hat a_k^\dagger\rangle;
  \end{equation}
for the normal ordering $\mathcal{P}=1$, one has
  \begin{equation}
    \langle \hat a_k^\dagger \hat a_k\rangle_{\mathcal{P}=1}=
    \langle \hat a_k^\dagger \hat a_k\rangle,
  \end{equation}
finally for antinormal ordering $\mathcal{P}=-1$,
  \begin{equation}
    \langle \hat a_k^\dagger \hat a_k\rangle_{\mathcal{P}=-1}=
    \langle \hat a_k \hat a_k^\dagger \rangle\;.
  \end{equation}
In general,
  \begin{equation}
    \label{eq:photoncounter}
    \langle \hat a_k^\dagger \hat a_k\rangle_{\mathcal{P}}=
    \langle \hat a_k^\dagger \hat a_k\rangle+\frac{1}{2}(1-\mathcal{P}),
  \end{equation}
  which is adopted below for the NN layer that returns the
  mean particle number. When not explicitly stated, we refer to symmetric ordering $\mathcal{P}=0$.

For use with the machine learning application programming interfaces as {\tt TensorFlow}, it is convenient to use real variables. We consider the  quadrature operators $\hat x_k$ and $\hat p_k$ defined by
\begin{equation}
\hat a_k=\frac{\hat{q}_k+\imath \hat{p}^\dagger_k}{\sqrt{2}}.
\end{equation}
The quadrature operators can be cast in operator vectors $\hmbfq$ and $\hmbfp$ with dimension
$n\times 1$,
\begin{equation}
  \hmbfq=\begin{pmatrix}\hq_0\\\hq_1\\\vdots\\\hq_{n-1}\end{pmatrix}\hspace{1cm}
  \hmbfp=\begin{pmatrix}\hp_0\\\hp_1\\\vdots\\\hp_{n-1}\end{pmatrix}.
  \end{equation}
In addition, we consider the real quantities
\begin{equation}\label{eq:pqdefinition}
    q_k=\frac{z_k-z_k^*}{\sqrt{2}\imath}\hspace{1cm}
    p_k=\frac{z_k+z_k^*}{\sqrt{2}},
  \end{equation}
  and we collect them in two real vectors $\mbfq$ and $\mbfp$ with dimension $1\times n$
  \begin{equation}
    \begin{array}{l}
  \mbfq=\begin{pmatrix}q_0 & q_1 &\cdots & q_{n-1}\end{pmatrix}\\
      \mbfp=\begin{pmatrix}p_0 & p_1 &\cdots & p_{n-1}\end{pmatrix}.
                                               \end{array}
  \end{equation}
Correspondingly, we have
\begin{equation}
\chi= \displaystyle\chi(\mbfq,\mbfp,\mathcal{P}=0)=\Tr [\rho \exp
    \left( \imath\, \mbfq\cdot \hat\mbfq+\imath\, \mbfp\cdot \hat\mbfp\right)]\text{.}
  \end{equation}
As  $\hat\mbfq$ and $\hat\mbfp$ are column vectors, and $\mbfq$ and $\mbfp$ are row vectors, we omit the dot product symbol and write
\begin{equation}
    \displaystyle\chi(\mbfq,\mbfp)=\Tr [\rho \exp
    \left( \imath\, \mbfq \,\hat\mbfq+\imath\, \mbfp \,\hat\mbfp\right)]\;.
\end{equation}
The notation can be simplified by defining the $1\times N$ real row vector
\begin{equation}
  \mbfx=\begin{pmatrix}q_0 & p_0 & q_1 & p_2 & \cdots q_n & p_n\end{pmatrix},
  \end{equation}
and we have
\begin{equation}
   \chi= \chi(\mbfx)=\Tr\left[\rho \exp
    \left( \imath\, \mbfx\, \hat\mbfR\right)\right]
    =\chi_R(\mbfx)+\imath \chi_I(\mbfx)
\end{equation}
being
\begin{equation}
  \begin{aligned}
    \mbfx\,\hat\mbfR&=\\&=q_0 \hat q_0+ p_0\hat p_0+q_1 \hat q_1+\ldots+ q_{n-1} \hat q_{n-1}+p_{n-1} \hat p_{n-1}=\\
    &=x_0\hat R_0+x_1\hat R_1+\ldots+x_{N-1}\hat R_{N-1}
    \end{aligned}
    \end{equation}
with the $N\times 1$ column operator vector
    \begin{equation}
      \label{eq:Rvector}
      \hat{\mbfR}=\begin{pmatrix}\hat q_0 \\
        \hat p_0\\ \hat q_1\\ \hat p_1\\\vdots\\\hat q_{n-1}\\ \hat p_{n-1}\end{pmatrix}\;.
    \end{equation}
Following the canonical commutation relations of the quadratures, in units with $\hbar=1$,
  \begin{equation}
    \begin{aligned}
      \left[\hat{q_j},\hat{q}_k\right]&=0\\
      \left[\hat{p_j},\hat{p}_k\right]&=0\\
      \left[\hat{q_j},\hat{p}_k\right]&=i\delta_{jk},
    \end{aligned}
    \end{equation}
  we have
  \begin{equation}\label{eq:commutatorR}
    [\hat{R}_p,\hat{R}_q]=\imath J_{pq},
  \end{equation}
  where we introduced the $N\times N$ symplectic matrix $\mbfJ$
  \begin{equation}
    \mbfJ=\bigoplus_{j} \mbfJ_1=\begin{pmatrix}
      0 & 1 & 0 & 0 & \cdots & 0 &0\\
      -1 & 0 & 0 & 0 & \cdots & 0 &0\\
      0 & 0 & 0 & 1 & \cdots & 0 &0\\
      0 & 0 & -1 & 0 & \cdots & 0 &0\\
      \vdots&\vdots&\vdots&\vdots& \ddots &\vdots&\vdots\\
      0 & 0 & 0 & 0 & \cdots & 0 &1\\
      0 & 0 & 0 & 0 & \cdots & -1 &0
    \end{pmatrix},
    \label{eq:symplecticmatrix}
  \end{equation}
  being  $\mbfJ_1=\big(\begin{smallmatrix} 0 &1 \\ -1& 0\end{smallmatrix}\big)$~\cite{X.2007}. In~\eqref{eq:symplecticmatrix}, the symbol $\bigoplus$ is the direct sum for block matrices with $j$ running in $0,1,\ldots\,n-1$. The matrix $\mbfJ$ is such that $\mbfJ^{-1}=\mbfJ^\top$, $\mbfJ^2=-{\bf 1}_N$, and $\mbfJ^\top \mbfJ={\bf 1}_N$ is the $N\times N$ identity matrix.

The expected value of $\hmbfR$ is determined by the derivatives of the characteristic function,
\begin{equation}\label{eq:chidefinition}
    \chi(\mbfx)=\Tr\left(\rho e^{\imath\, \mbfx\,\hat {\bf R}}\right)\;.
\end{equation}
One has
\begin{equation}
\langle \hat{R}_j\rangle =\Tr[\rho\hat{R}_j]= -\imath \left.\frac{\partial \chi}{\partial x_j}\right\vert_{\mbfx =0}=\left.\frac{\partial \chi_I}{\partial x_j}\right\vert_{\mbfx=0}
\label{eq:derchiI}
\end{equation}
or
\begin{equation}
  \langle \hat{\mbfR}\rangle =\Tr[\rho\hat{\mbfR}]=
  -\imath \left.\nabla \chi\right\vert_{\mbfx =0}=\left.\nabla \chi_I\right\vert_{\mbfx=0}\;.
\label{eq:derchiI}
\end{equation}
As the components of ${\hat {\bf R}}$ are self-adjoint, their mean value is real,
and it is given by the derivative of the imaginary part $\chi_I$ at $\mbfx=0$.
We also have
\begin{equation}
  \begin{aligned}
    &-\frac{\partial^2}{\partial z_j \partial z_j^*}\chi=\\
    &=-\left(  \frac{\partial}{\partial q_j}\frac{\partial q_j}{\partial z_j}
    + \frac{\partial}{\partial p_j}\frac{\partial p_j}{\partial z_j}
  \right)
\left(  \frac{\partial}{\partial q_j}\frac{\partial q_j}{\partial z^*_j}
    + \frac{\partial}{\partial p_j}\frac{\partial p_j}{\partial z^*_j}
  \right)
    \chi\\&=
-\frac{1}{2} \left(  \frac{\partial^2}{\partial q_j^2}
    + \frac{\partial^2}{\partial p_j^2}
  \right)
  \chi,
  \end{aligned}
\end{equation}
after using~\eqref{eq:pqdefinition}. Thus, we evaluate the expected number of photons by the second derivatives of $\chi_R$
\begin{equation}
  \label{eq:photoncounter1}
  \displaystyle\langle \hat{a}^\dagger_j \hat{a_j} \rangle_{\mathcal{P}=0}=
\left. -\frac{1}{2} \left(  \frac{\partial^2}{\partial x_{2j}^2}
    + \frac{\partial^2}{\partial x_{2j+1}^2}
  \right)    \chi_R\right\vert_{\mbfx=0}
  \end{equation}
 with $j=0,1,2,\ldots,n-1$.

For the expected value of total number of particles $\hat{\mathcal{N}}$, we have\begin{equation}
  \displaystyle \langle\hat{\mathcal{N}}\rangle_{\mathcal{P}=0}=\left.-\frac{1}{2}\nabla^2\chi_R\right\vert_{\mbfx=0}
  \label{eq:N}
  \end{equation}
  the Laplacian being computed with respect to all the components of $\mbfx$.
  Equation~(\ref{eq:N}) refers to symmetric ordering with $\mathcal{P}=0$.
  For normal ordering, we use \eqref{eq:photoncounter} as
  \begin{equation}
    \langle \ha_k^\dagger \ha_k\rangle=\langle \ha_k^\dagger \ha_k\rangle_{\mathcal{P}=1} =\langle \ha_k^\dagger \ha_k\rangle_{\mathcal{P}=0}-\frac{1}{2}\;,
  \end{equation}
  and we get
\begin{equation}
\begin{array}{ll}
  \displaystyle \langle\hat{\mathcal{N}}\rangle&=
  \displaystyle \langle\hat{\mathcal{N}}\rangle_{\mathcal{P}=1}=\displaystyle \langle\hat{\mathcal{N}}\rangle_{\mathcal{P}=0}-\frac{n}{2}\\[2pt]&=\left.-\frac{1}{2}\nabla^2\chi_R\right\vert_{\mbfx=0}
  -\frac{n}{2}\\[2pt]&=-\frac{1}{2}\left.\left(\nabla^2+n\right)\chi\right|_{\mbfx=0}.
  \end{array}
\end{equation}
%%%%%%%%%%%%%%%%%%%%%%%%%%%%%%%%%
\section{Gaussian states}
For a Gaussian state,
\begin{equation}
  \chi(\mbfx)=\exp(-\frac{1}{4}\mbfx\, \mbfg\,\mbfx^\top +\imath\, \mbfx\, \mbfd)
  \label{eq:characteristicGaussian}
\end{equation}
with $\mbfg$ the $N\times N$ covariance matrix,
and $\mbfd$ the $N\times 1$ displacement column vector.
Correspondingly, being ${\mbfx}$ a row vector,
\begin{equation}
  \begin{array}{l}
    \chi_R(\mbfx)\displaystyle =e^{-\frac{\mbfx \mbfg \mbfx^\top}{4}}\cos(\mbfx\mbfd)\\
     \chi_I(\mbfx)\displaystyle=e^{-\frac{\mbfx \mbfg \mbfx^\top}{4}}\sin(\mbfx\mbfd)
    \end{array}
\end{equation}
From Eq.~(\ref{eq:derchiI}), we have for $\mbfd$, with $q=0,1,\ldots,N-1$.
\begin{equation}
\langle \hat{R}_q\rangle =\Tr[\hat{\rho} \hat{R}_q]=d_q= \left.\frac{\partial \chi_I}{\partial x_q}\right\vert_{\mbfx =0}
\label{eq:chiIderivative}
\end{equation}
or
\begin{equation}
    \langle \mbfR \rangle =\mbfd =\left.\nabla \chi_I\right\vert_{\mbfx=0}\;.
\end{equation}
Hence, the displacement $d_q$ is the first moment of the Gaussian $\chi$.
The derivatives of the characteristic function can be expressed by the
displacement and covariance matrix $d_q$ and $g_{pq}$.
This speeds up computing the derivatives of $\chi$,
and the observable quantities.

For the particle number, we have
\begin{equation}
  \begin{aligned}
    \langle \hat{a}^\dagger_j \hat{a_j} \rangle_{\mathcal{P}=0}&=\displaystyle
    \left.-\frac{1}{2} \left(  \frac{\partial^2}{\partial x_{2j}^2}
      + \frac{\partial^2}{\partial x_{2j+1}^2}
    \right)\right|_{\mbfx=0}  \chi_R \\&= \frac{g_{2j,2j}+g_{2j+1,2j+1}}{4}+
    \frac{d_{2j}^2+d_{2j+1}^2}{2}
  \end{aligned}
\end{equation}
and
\begin{equation}
  \begin{aligned}
    \langle \hat{a}^\dagger_j \hat{a_j} \rangle&=\displaystyle
    -\frac{1}{2} \left(  \frac{\partial^2}{\partial x_{2j}^2}
    + \frac{\partial^2}{\partial x_{2j+1}^2}
    \right)    \chi_R -\frac{1}{2}\\&=\frac{g_{2j,2j}+g_{2j+1,2j+1}}{4}+
    \frac{d_{2j}^2+d_{2j+1}^2}{2}-\frac{1}{2}
  \end{aligned}
    \label{eq:avgnumberj}
\end{equation}
while the second derivatives of $\chi_I$ at $\mbfx=0$ are vanishing,
and we used~\eqref{eq:photoncounter}.
%\footnote{The detailed derivation of~\eqref{eq:avgnumberj} is in Section~\ref{sec:Proof1}.}
The covariance matrix $\mbfg$ is determined by the second moments
\begin{equation}\label{eq:covariance1}
  g_{pq}=
  \langle\{\hat{R}_p,\hat{R}_q\}\rangle-2\langle{\hat{R}_p}\rangle\langle{\hat{R}_q}\rangle
\end{equation}
with
\begin{equation}
  \{\hat{R}_p, \hat{R}_q\}=\hat{R}_p\hat{R}_q +\hat{R}_q \hat{R}_p.
\end{equation}
We also have after \eqref{eq:commutatorR}
\begin{equation}\label{eq:covariance2}
  \begin{aligned}
    g_{pq}&=\langle \hat{R}_p\hat{R}_q\rangle+\langle \hat{R}_q\hat{R}_p\rangle-2\langle{R_p}\rangle\langle{R_q}\rangle\\
    &= \langle \hat{R}_p\hat{R}_q\rangle+\langle \hat{R}_q\hat{R}_p\rangle-2\langle{R_p}\rangle\langle{R_q}\rangle\\
    &= 2\langle \hat{R}_p\hat{R}_q\rangle-\imath J_{pq}-2\langle{R_p}\rangle\langle{R_q}\rangle=\\
    &=2\langle \left(\hat{R}_p-d_p\right)\left(\hat{R}_q-d_q\right)\rangle-\imath J_{pq}\;.
    \end{aligned}
\end{equation}
\subsection{Vacuum state}
Vacuum states are Gaussian state with $\mbfd = {\bf 0}$ and $\mbfg={\bf 1}_{N}$.
From \eqref{eq:avgnumberj}, we get
\begin{equation}
    \langle \hat{a}^\dagger_j \hat{a_j} \rangle_{\mathcal{P}=0}=\frac{1}{2}\;.
  \end{equation}
  Correspondingly,
  \begin{equation}
        \langle \hat{a}^\dagger_j \hat{a_j} \rangle=\langle \hat{a}^\dagger_j \hat{a_j} \rangle_{\mathcal{P}=0}-\frac{1}{2}=0.
  \end{equation}
We have $\langle  \hat{\mbfR}\rangle=0$, and for the average photon number after Eq.~(\ref{eq:avgnumberj})
\begin{equation}
  \langle  \mathcal{N}\rangle=\sum_{j=0}^{n-1}\langle  \hat{a}^\dagger_j \hat{a}_j\rangle= 0\;.
  \end{equation}
\subsection{Coherent state}\label{sec:coherentstate}
A coherent state has a non-vanishing displacement vector $\mbfd\neq {\bf 0}$ and $\mbfg={\bf 1}_N$.
For a single mode $|\alpha\rangle$, with $n=1$ and $N=2$,
\begin{equation}
  \begin{array}{l}
    d_0=\sqrt{2}\Re\left({\alpha}\right)\displaystyle=\frac{\alpha+\alpha^*}{\sqrt{2}}\\
    d_1=\sqrt{2}\Im\left({\alpha}\right)\displaystyle=\frac{\alpha-\alpha^*}{\imath\sqrt{2}}\text{.}
  \end{array}
\end{equation}
From~\eqref{eq:avgnumberj}, we have ($j=0$)
\begin{equation}
\langle \hat{a}^\dagger_j \hat{a_j} \rangle=\langle \hat{a}^\dagger_j \hat{a_j} \rangle_{\mathcal{P}=0}-\frac{1}{2}=|\alpha|^2= \frac{d_0^2+d_1^2}{2}.
  \end{equation}
  For $n$ modes, $j=0,1,\ldots,n-1$ we have
  \begin{equation}
\langle \hat{a}^\dagger_j \hat{a_j} \rangle= \frac{d_{2j}^2+d_{2j+1}^2}{2},
\end{equation}
and
\begin{equation}
  \langle  \mathcal{N}\rangle=\sum_{j=0}^{n-1}\langle  \hat{a}^\dagger_j \hat{a}_j\rangle= \sum_{q=0}^{N-1}{\frac{d^2_q}{2}}.
\end{equation}
\subsection{Covariance matrix}
Given $\chi$, one can compute the covariance matrix and the displacement operator by derivation.
$\mbfd$ can be computed by the first derivatives, as in \eqref{eq:derchiI}
\begin{equation}\label{eq:derchiI1}
d_q=\langle \hat{R}_q\rangle =\Tr[\hat{\rho} \hat{R}_q]= \left.\frac{\partial \chi_I}{\partial x_q}\right\vert_{\mbfx =0}.
\end{equation}
Writing the characteristic function in terms of the components of $\mbfx$
\begin{equation}
    \chi(\mbfx)=\exp\left(-\frac{1}{4}\sum_{pq} g_{pq} x_p x_q+\imath \sum_{p} x_p d_p \right)\;.
\label{eq:proofCOV0}
\end{equation}
with $p,q=0,1,...,N-1$, we have for the first derivative, by using the symmetry $g_{pq}=g_{qp}$,
\begin{equation}
\frac{\partial{\chi}}{{\partial x_m}}=
\left(-\frac{1}{2}\sum_p g_{mp}x_p +\imath d_m \right)\chi(\mbfx)
\label{eq:proofCOV1}
\end{equation}
with $m=0,1,...,N-1$.
Evaluating Eq.~(\ref{eq:proofCOV1}) at $\mbfx=0$, we obtain as above,
being $\chi({\bf 0})=1$,
\begin{equation}
\left.\frac{\partial{\chi}}{{\partial x_m}}\right|_{\mbfx=0}=
\imath \left.\frac{\partial{\chi_I}}{{\partial x_m}}\right|_{\mbfx=0}=\imath d_m\;.
\label{eq:proofCOV2}
\end{equation}
For the second derivatives, we have
\begin{equation}
\begin{array}{l}
\displaystyle\left.\frac{\partial^2{\chi_R}}{{\partial x_m}{\partial x_n}}\right|
_{\mbfx=0}=
-\frac{1}{2}g_{mn}-d_m d_n\\
\displaystyle\left.\frac{\partial^2{\chi_I}}{{\partial x_m}{\partial x_n}}\right|
_{\mbfx=0}\textbf{}=
0\;.\\
\end{array}
\label{eq:proofCOV4}
\end{equation}
The covariance matrix is given by
\begin{equation}\label{eq:dercov}
  \begin{aligned}
  g_{pq}=
  -2\left.\frac{\partial^2 \chi_R}{\partial x_p \partial x_q}\right|_{\mbfx=0}-2d_p d_q\\=
    \left.-2\frac{\partial^2 \chi_R}{\partial x_p \partial x_q}-2\frac{\partial \chi_I}{\partial x_p}\frac{\partial \chi_I}{\partial x_q}\right|_{\mbfx=0}
    \end{aligned}
 \end{equation}
 Equation~(\ref{eq:dercov}) is helpful to compute the covariance matrix after the model is trained to determine various features as, e.g., the degree of entanglement.
%%%%%%%%%%%%%%%%%%%%%%%%%%%%%%%%%%%%%%%%%%%%%%%%%%%%%%
\section{Gates as linear layers}\label{chp:lineartrans}
Gates represented by unitary operations are of relevance
in many applications. We are interested to those gates such that the new annihilation operators are expressed as a linear combination of the input operators. If the new operator is
\begin{equation}
  \hat{\widetilde{\mbfa}}=\hat U^{\dagger}\hat{\mbfa}\hat U\;,
\end{equation}
in term of the density matrix, the transformation reads~\cite{BarnettBook}
\begin{equation}
    \tilde \rho =\hat U\rho\, \hat U^{\dagger}.
\end{equation}

We start considering a transformation such that~\cite{BarnettBook}
\begin{equation}
\displaystyle    \hat{\widetilde{\mbfa}} =\mbfU \hat{\mbfa}.
    \label{eq:lineartransform}
\end{equation}
with ${\mbfU}$ a $n\times n$ complex matrix, with $\hat{\mbfa}$ and
$\hat{\widetilde{\mbfa}}$ column vectors of operators with dimension $n\times 1$.
In a later section, we will consider the more general case with
\begin{equation}
\hat{\widetilde{\mbfa}}=\mbfU \hat{\mbfa}+\mbfW \hat{\mbfa}^\dagger.
\end{equation}
For the moment, we have $\mbfW=0$.

The transformation in terms of the canonical variables read~\cite{X.2007}
\begin{equation}
    \hat{\widetilde{\bf R}}=\hat U^\dagger \hat{{\bf R}} \hat U=
    \mbfM\, \hat{{\bf R}}+\mbfd'
    \label{eq:Rtransform}
    \end{equation}

Linear transformations transform Gaussian states into Gaussian states~\cite{X.2007}.
A Gaussian state with covariance matrix
$\mbfg$ and displacement vector $\mbfd$,
turns into a new Gaussian state with covariance matrix
\begin{equation}
  \label{eq:transformg}
    \widetilde \mbfg = \mbfM\, {\mbfg}\,\mbfM^\top
\end{equation}
and displacement vector
\begin{equation}
  \label{eq:transformd}
\widetilde{\mbfd}=\mbfM\,\mbfd+\mbfd'\text{.}
\end{equation}

If the transformation due to $\mbfU$ is unitary, the matrix $\mbfM$ is
symplectic, and satisfies
\begin{equation}
  \mbfM \mbfJ \mbfM^\top=\mbfJ
\end{equation}
with $\mbfJ$ given in~\eqref{eq:symplecticmatrix}.
The inverse of $\mbfM$ is found as
\begin{equation}
\mbfM^{-1}=\mbfJ \mbfM^\top\mbfJ^{\top}\;.
\end{equation}
%%%%%%%%%%%%%%%%%%%%%%%%%%%%%%%%%%%%%%%%%%%%%%%%%%%%%%%%
%\section{The $\mbfU$ and $\mbfM$ matrices}\label{sec:RPRQ}
It is instructive to deepen the link between $\mbfU$ and $\mbfM$ for later use.
We consider the $n\times 1$ vector of the annihilation operators $\hat {\mbfa}$,
\begin{equation}
  \hat{\mbfa}=\begin{pmatrix}\hat{a}_0\\
  \hat{a}_1\\\vdots\\\hat{a}_{n-1}\end{pmatrix}
  \end{equation}
  and the corresponding $n\times 1$ vectors of positions $\hat\mbfq$ and
  momentum vector $\hat\mbfp$,
\begin{equation}
  \hat{\mbfq}=\begin{pmatrix}\hat{q}_0\\
  \hat{q}_1\\\vdots\\\hat{q}_{n-1}\end{pmatrix}
\hspace{1cm}
  \hat{\mbfp}=\begin{pmatrix}\hat{p}_0\\
  \hat{p}_1\\\vdots\\\hat{p}_{n-1}\end{pmatrix}.
\end{equation}

We build the $\hat {\bf R}$ vector in~\eqref{eq:Rvector} by using auxiliary rectangular matrices $\mbfRq$ and $\mbfRp$ as follows
\begin{equation}
    \hat\mbfR = \mbfRq \hat \mbfq+\mbfRp \hat \mbfp.
\end{equation}
$\mbfRq$ and $\mbfRp$ are matrices
with size $N\times n$.

For $N=4$, we have
\begin{equation}
\mbfRq=
\begin{pmatrix}
1 & 0\\ 0 & 0 \\ 0 & 1 \\ 0 & 0
\end{pmatrix}
\end{equation}
and
\begin{equation}
\mbfRp=
\begin{pmatrix}
0 & 0\\ 1 & 0 \\ 0 & 0 \\ 0 & 1
\end{pmatrix},
\end{equation}
such that
\begin{equation}
\mbfRq \hat{{\bf q}}=
\begin{pmatrix}
\hat q_0\\ 0 \\ \hat q_1 \\ 0
\end{pmatrix},
\end{equation}
and
\begin{equation}
\mbfRp \hat{{\bf p}}=
\begin{pmatrix}
0\\ \hat p_0 \\ 0 \\ \hat p_1
\end{pmatrix},
\end{equation}
and hence
\begin{equation}
{\bf \hat R}=
\mbfRq \hat{{\bf q}}+
\mbfRp \hat{\mbfp}=
\begin{pmatrix}
\hat q_0 \\ \hat p_0 \\ \hat q_1 \\ \hat p_1,
\end{pmatrix}\;,
\end{equation}
being
\begin{equation}
\hat{\mbfq}=
\begin{pmatrix}
\hat q_0 \\ \hat q_1
\end{pmatrix},
\end{equation}
and
\begin{equation}
\hat{\mbfp}=
\begin{pmatrix}
\hat p_0 \\ \hat p_1
\end{pmatrix}.
\end{equation}
From the previous expressions, one has
\begin{equation}
\begin{array}{l}
   \hat{\mbfq}= \mbfRq^\top \hat{\mbfR}\\
   \hat{\mbfp} = \mbfRp^\top \hat\mbfR
    \end{array}
\end{equation}
and the rectangular matrices $\mbfRq$ and $\mbfRp$ satisfy
\begin{equation}
\begin{array}{l}
    \mbfRq^\top \mbfRq = {\bf 1}_{n}\\
    \mbfRq^\top \mbfRp = {\bf 0}_{n}\\
    \mbfRp^\top \mbfRp = {\bf 1}_{n}\\
    \mbfRp^\top \mbfRq = {\bf 0}_{n}\\
    \mbfRq \mbfRq^\top+\mbfRp \mbfRp^\top = {\bf 1}_{N}\\
    \mbfJ \mbfRq = -\mbfRp\\
    \mbfJ \mbfRp = \mbfRq\\
    \mbfRq \mbfRp^\top-\mbfRp \mbfRq^\top = \mbfJ
    \end{array}\;.\label{eq:Rpropterties}
  \end{equation}
By the matrices $\mbfR_q$ and $\mbfR_p$, we can write the $\hat\mbfa$ vector as follows
\begin{equation}
    \hat\mbfa = \frac{\hat\mbfq+\imath \hat\mbfp}{\sqrt{2}}=
    \frac{\mbfRq^\top+\imath \mbfRp^\top}{\sqrt{2}}\hat{\bf R}.
\end{equation}
Letting  $\mbfU=\mbfU_R+\imath\mbfU_{I}$ with real part $\mbfU_R$ and imaginary part $\mbfU_I$;
and $\mbfU^*=\mbfU_R-\imath\mbfU_{I}$.
Correspondingly
\begin{equation}
\begin{array}{l}
\hat{\widetilde\mbfq} = \mbfU_R \hat\mbfq - \mbfU_I \hat\mbfp\\
\hat{\widetilde\mbfp} = \mbfU_I \hat\mbfq + \mbfU_R \hat\mbfp
\end{array}
\label{eq:qptransform}
\end{equation}
As $\hat\mbfR$ transforms in $\hat{\widetilde\mbfR}$, we have
\begin{equation}
    \hat{\widetilde\mbfR}=\mbfRq\hat{\widetilde\mbfq}+
\mbfRp\hat{\widetilde\mbfp}=\mbfM\hat{\mbfR}=\left(\mbfM_1+\mbfM_2\right)\hat{\mbfR}
\end{equation}
with $\mbfM_1$, $\mbfM_2$, and $\mbfM$ real matrices such that
\begin{equation}
  \begin{array}{l}
    \mbfM=\mbfM_1+\mbfM_2\\
    \mbfM_1 = \mbfRq \mbfU_R\mbfRq^\top+\mbfRp \mbfU_R\mbfRp^\top\\
    \mbfM_2 = \mbfRp \mbfU_I\mbfRq^\top-\mbfRq \mbfU_I\mbfRp^\top
  \end{array}
\label{eq:symplecticmain}
\end{equation}
Equations~(\ref{eq:symplecticmain}) are used in defining
linear gates starting from their matrix $\mbfU$.
\section{Gaussian states as neural networks}
NNs are mathematical tools to approximate unknown functions of many variables. NNs depend on many parameters that can be fine tuned to improve fitting a target model.
Gaussian states correspond to NNs with a Gaussian activation function, which is often adopted, for example, in {\it kernel machines}~\cite{SchuldPetruccioneBook}.
\begin{figure}
  \centering
  \includegraphics{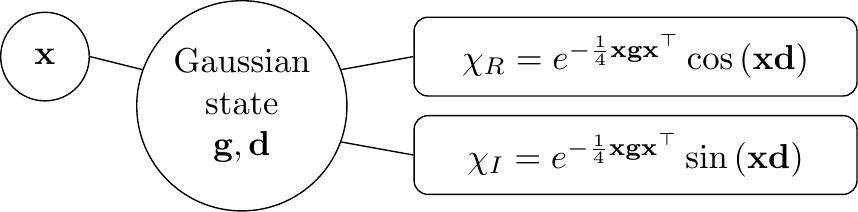}
\caption{Double-head model for a layer representing a Gaussian.\label{fig:chpGNNfig1}}
\end{figure}
%%%%%%%%%%%%%%%%%%%%%%%%%%%%%%%%%%%%%%%%%%%%%
We implement a Gaussian density matrix by
a ``multi-head model'' that is a model, which
has a computational backbone that processes the inputs and multiple output layers to return different quantities. We start considering as ``heads'' the real and imaginary parts $\chi_R$ and $\chi_I$.

Figure~\ref{fig:chpGNNfig1} shows the NN model for a Gaussian state with each layer having one input $\mbfx$ and two outputs (the ``heads'') $\chi_R$ and $\chi_I$. The input is the state row vector $\mbfx$, and is the input layer. The outputs $\chi_R$ and $\chi_I$ are the output layers.
The inner layer denoted ``Gaussian state,'' computes the Gaussian characteristic function.
The parameters of this layer, namely the covariance matrix $\mbfg$ and the displacement $\mbfd$ are also indicated.

For a Gaussian state with covariance matrix ${\bf g}$ and displacement vector ${\bf d}$ we have
\begin{equation}
\begin{array}{ll}
\chi_R(\mbfx)&=\exp(-\frac{1}{4}\mbfx {\bf g} \mbfx^\top)
\cos(\mbfx \mbfd)\\[2pt]
  \chi_I(\mbfx)&=\exp(-\frac{1}{4}\mbfx {\bf g} \mbfx^\top)
\sin(\mbfx \mbfd).
\end{array}
\label{eq:gaussianchirchi}
\end{equation}
However, it is convenient to consider a more general model that includes a further input vector (the ``bias'').

\begin{figure}
  \centering
  \includegraphics{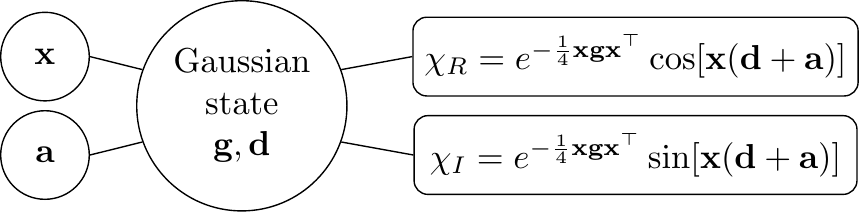}
\caption{Double-head model representing a Gaussian state with a bias.\label{fig:chpGNNfig2}}
\end{figure}

We generalize the Gaussian layer to include a further bias input ${\bf a}$ with the same dimensions of $\mbfd$, in addition to ${\mbfx}$, such that the output is
\begin{equation}
  \chi(x)=e^{-\frac{1}{4}\mbfx {\bf g} \mbfx^T}e^{\imath(\mbfd+\mbfa)},
  \label{eq:chigen}
\end{equation}
Here $\mbfa$ is a bias in the displacement, which will be useful for cascading layers as detailed below.
Equation~(\ref{eq:chigen}) corresponds to the two heads of the model returning
\begin{equation}
\begin{array}{ll}
\chi_R(\mbfx)&=\exp(-\frac{1}{4}\mbfx {\bf g} \mbfx^T)
\cos[\mbfx (\mbfd+{\bf a})]\\[2pt]
\chi_I(\mbfx)&=\exp(-\frac{1}{4}\mbfx {\bf g} \mbfx^T)
\sin[\mbfx (\mbfd+{\bf a})].
\end{array}
\end{equation}
Figure~\ref{fig:chpGNNfig2} shows a graphical representation of the generalized Gaussian NN with bias.
\section{Pullback}\label{chp:pullback}
%\section{Introduction}
We setup our NN representation of the density matrix as a layered sequence of gates. Each gate is a unitary operator that acts on the density matrix, and transform the latter into a new state. To detail how it works, we start considering the transformation of the characteristic function under unitary operations.

We consider a linear transformation by the operator $\hat{U}$.
The operators and the density matrix changes as follows
\begin{equation}
  \begin{array}{l}
    \hat{\widetilde{\mbfa}}=\hat{U}^\dagger \hat{\mbfa}\hat{U}\\
    \tilde \rho =\hat U\rho\, \hat U^{\dagger}.
    \end{array}
\end{equation}
Correspondingly, for the transformed characteristic functions $\tilde\chi$, we have
\begin{equation}
  \begin{aligned}
  \tilde\chi&=  \Tr\left[\widetilde{\rho} e^{\imath \mbfx \hat{\mbfR}}\right]
  =\Tr\left[\hat{U}\rho\hat{U}^\dagger e^{\imath \mbfx \hat{\mbfR}}\right]=
  \Tr\left[\rho\hat{U}^{\dagger} e^{\imath \mbfx \hat{\mbfR}}\hat{U}\right]
  \\
 &= \Tr\left[\rho e^{\imath \mbfx \hat{U}^\dagger\hat{\mbfR}\hat{U}}\right]=
  \Tr\left[\rho e^{\imath \mbfx \hat{\widetilde{\mbfR}}}\right].
  \end{aligned}
\end{equation}
Recalling Eq.~(\ref{eq:Rtransform})
\begin{equation}
  \label{eq:Rtransform:2}
  \hat{\widetilde{\mbfR}}=\mbfM\mbfR+\mbfd',
\end{equation}
we remark that the linear transformation is determined by the symplectic
matrix $\mbfM$ and the displacement $\mbfd'$.
We write $\tilde\chi$ in terms of $\mbfM$ and $\mbfd'$, by using~\eqref{eq:Rtransform:2},
  \begin{equation}
    \tilde \chi=\Tr\left[\rho e^{\imath \mbfx \mbfM \hat{\mbfR}}\right]e^{\imath \mbfx \mbfd'}.
    \label{eq:chitilde}
  \end{equation}
  On the other hand, the expression for $\chi(\mbfx)$ is the following
  \begin{equation}
    \chi=\Tr\left[\rho e^{\imath \mbfx \hat{\mbfR}}\right].
    \label{eq:chi}
  \end{equation}
  Thus we have
  \begin{equation}
    \tilde\chi(\mbfx)=\chi(\mbfx\mbfM)e^{\imath \mbfx \mbfd'}=\chi(\mbfx\mbfM)e^{\imath (\mbfx\mbfM) (\mbfM^{-1}\mbfd')}.
    \label{eq:chitilde}
  \end{equation}
From~\eqref{eq:chitilde}, we see that the modified characteristic function depends on the modified input vector $\mbfx\mbfM$ and has a modified displacement $\mbfM^{-1}\mbfd'$.

\begin{figure}
    \centering
  \includegraphics[width=0.33\textwidth]{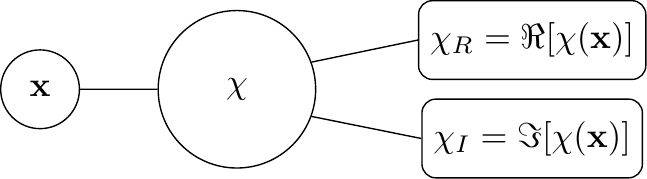}
  \caption{Graphical rapresentation of the model for the characteristic function.\label{fig:chpPBKfig1}}
  \end{figure}

\begin{figure}
    \centering
  \includegraphics[width=0.33\textwidth]{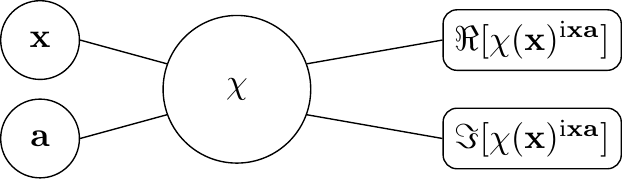}
  \caption{Graphical rapresentation of the generalized model for the characteristic function
      with bias $\mbfa$, which is adopted in the linear transformations.\label{fig:chpPBKfig2}}
  \end{figure}

\begin{figure}
    \centering
  \includegraphics[width=0.45\textwidth]{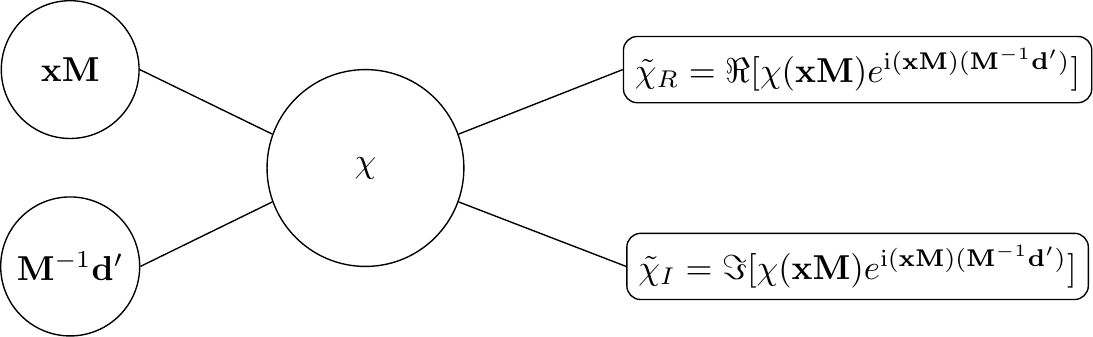}
  \caption{Graphical representation of the transformed characteristic function in terms of the original characteristic function with modified input and bias.\label{fig:chpPBKfig3}}
  \end{figure}

Introducing the new input vector $\mbfy=\mbfx\mbfM$ and the bias $\mbfa=\mbfM^{-1}\mbfd'$,
we can express the transformed characteristic function $\tilde\chi$ in terms of the original $\chi$ as follows
  \begin{equation}
    \label{eq:chitildetr}
    \tilde\chi(\mbfx)=\chi(\mbfy)e^{\imath \mbfy \mbfa}.
  \end{equation}
  We give a graphical representation of the transformations in Eq.~(\ref{eq:chitildetr}).
  We start considering a characteristic function with input vector $\mbfx$ and multi-headed outputs $\chi_R$ and $\chi_I$,  as in figure~\ref{fig:chpPBKfig1}.
  Then we generalize the model by introducing the bias vector $\mbfa$ as in the previous section and shown in Fig.~\ref{fig:chpPBKfig2}. Fig.~\ref{fig:chpPBKfig3} represents the transformation in equation~\eqref{eq:chitildetr}.
%%%%%%%%%%%%%%%%%%%%%%%%%%%%%%%%%%%%%%%%%%%%%
%%%%%%%%%%%%%%%%%%%%%%%%%%%%%%%%%%%%%%%%%%%%%
%%%%%%%%%%%%%%%%%%%%%%%%%%%%%%%%%%%%%%%%%%%%%
%%%%%%%%%%%%%%%%%%%%%%%%%%%%%%%%%%%%%%%%%%%%%
%%%%%%%%%%%%%%%%%%%%%%%%%%%%%%%%%%%%%%%%%%%%%
  %%%%%%%%%%%%%%%%%%%%%%%%%%%%%%%%%%%%%%%%%%%%%

\begin{figure}
    \centering
  \includegraphics{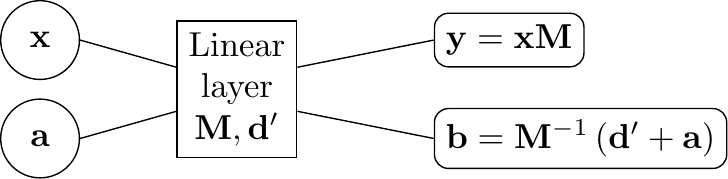}
  \caption{Linear layer to transform the input variables to the $\chi$ layer.\label{fig:chpPBKfig4}}
  \end{figure}

We introduce a new layer, which we call ``linear layer'', having as parameters the symplectic matrix $\mbfM$ and the displacement $\mbfd'$.
 The linear layer is shown in Fig.~\ref{fig:chpPBKfig4}, and it has two inputs $\mbfx$ and $\mbfa$, and two outputs: $\mbfy=\mbfx\mbfM$, a row vector with the same size of $\mbfx$, and a new displacement $\mbfb=\mbfM^{-1}(\mbfd'+\mbfa)$ a column vector with the size of $\mbfd$. The linear layer will enable us to represent many different linear transformations,  as squeezing or Glauber operators.

\begin{figure}
    \centering
  \includegraphics[width=0.33\textwidth]{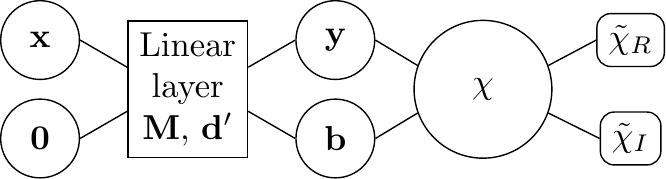}
    \caption{Graphical representation of a linear transformation as a pullback
      of a linear layer from the original characteristic function.
      The resulting model corresponds the transformed characteristic function in Fig.~\ref{fig:chpPBKfig3}. Note that here $\mbfa=0$.\label{fig:chpPBKfig5}}
  \end{figure}

\begin{figure}
    \centering
    \includegraphics{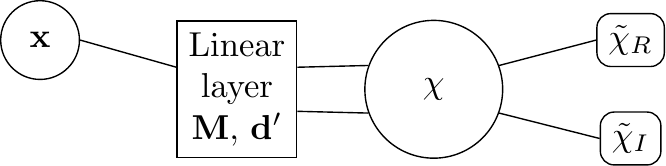}
    \caption{Simplified model of Fig.~\ref{fig:chpPBKfig5}, omitting the internal variables $\mbfy$ and $\mbfb$ and the zero input bias $\mbfa$.\label{fig:chpPBKfig6}}
  \end{figure}

By using the linear layer, we represent the transformation by cascading two layers.
 The resulting model is Fig.~\ref{fig:chpPBKfig5}. The model can be described as the pullback of the  linear layer from the characteristic function layer. The matrix $\mbfM$ first acts on the input $\mbfx$, and then the function $\chi$ is evaluated on the vector $\mbfy$.
A simplified diagram is given in Fig.~\ref{fig:chpPBKfig6}. This representation will be useful when considering multiple transformations.
%%%%%%%%%%%%%%%%%%%%%%%%%%%%%%%%%%%%%%%%%%%%%
%%%%%%%%%%%%%%%%%%%%%%%%%%%%%%%%%%%%%%%%%%%%%
%%%%%%%%%%%%%%%%%%%%%%%%%%%%%%%%%%%%%%%%%%%%%
%%%%%%%%%%%%%%%%%%%%%%%%%%%%%%%%%%%%%%%%%%%%%
%%%%%%%%%%%%%%%%%%%%%%%%%%%%%%%%%%%%%%%%%%%%%
%%%%%%%%%%%%%%%%%%%%%%%%%%%%%%%%%%%%%%%%%%%%%
%\section{Pullback of Gaussian states}

\begin{figure}
  \centering
  \includegraphics[width=0.45\textwidth]{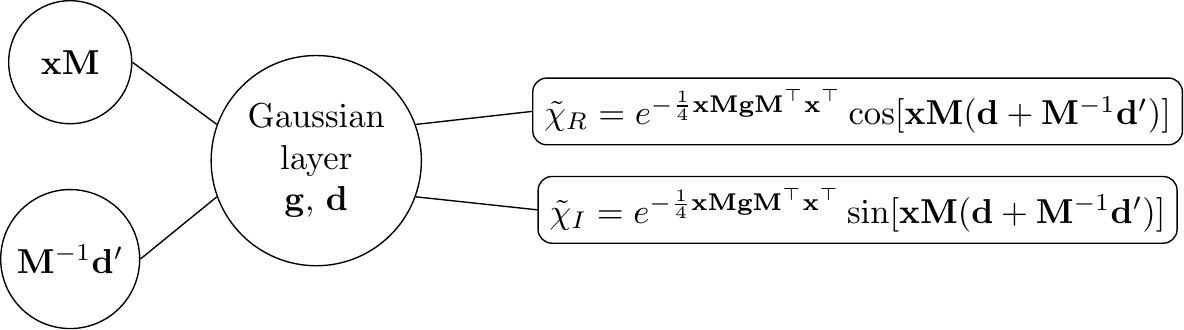}
  \caption{Single-layer multiheaded model for a linear transformation of a Gaussian state.\label{fig:chpPBKfig7}}
  \end{figure}

\begin{figure}
    \centering
\includegraphics{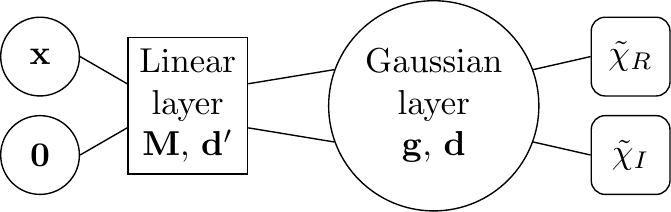}
  \caption{Two-layer multiheaded model equivalent to Fig.~\ref{fig:chpPBKfig7}.\label{fig:chpPBKfig8}}
  \end{figure}

For a Gaussian state [see Eq.~(\ref{eq:characteristicGaussian})]
the transformed characteristic function reads
  \begin{equation}
    \tilde\chi(\mbfx)=e^{-\frac{1}{4}\mbfx \mbfM \mbfg \mbfM^\top \mbfx^\top}e^{\imath \mbfx \mbfM(\mbfd+\mbfM^{-1} \mbfd')}.
    \label{eq:chitilde1}
  \end{equation}
Equation~\eqref{eq:chitilde1}  is graphically represented in Fig.~\ref{fig:chpPBKfig7}, by using the Gaussian layer in Fig.~\ref{fig:chpGNNfig2}.  This representation is equivalent to Fig.~\ref{fig:chpPBKfig8},  by the linear layer, whose action is detailed in Fig.~\ref{fig:chpPBKfig5}.
%%%%%%%%%%%%%%%%%%%%%%%%%%%%%%%%%%%%%%%%%%%%%
%%%%%%%%%%%%%%%%%%%%%%%%%%%%%%%%%%%%%%%%%%%%%
%%%%%%%%%%%%%%%%%%%%%%%%%%%%%%%%%%%%%%%%%%%%%
%%%%%%%%%%%%%%%%%%%%%%%%%%%%%%%%%%%%%%%%%%%%%
As shown in Fig.~\ref{fig:chpPBKfig8} linear transformation on the density matrix is actually a transformation of the variable $\mbfx$. This implies that the linear transformation
can be represent first by a linear gate, followed by the Gaussian gate.
This may be described as ``pulling back" the linear operation before the Gaussian gate.
By using these two models, the linear transformation of the Gaussian state is a cascade
of a linear pullback layer and a Gaussian state layer, as shown in figure \ref{fig:chpPBKfig8}.
%%%%%%%%%%%%%%%%%%%%%%%%%%%%%%%%%%%%%%%% Pullback cascading
\section{Pullback cascading}\label{sec:cascading}
\begin{figure*}
  \centering
\includegraphics[width=0.8\textwidth]{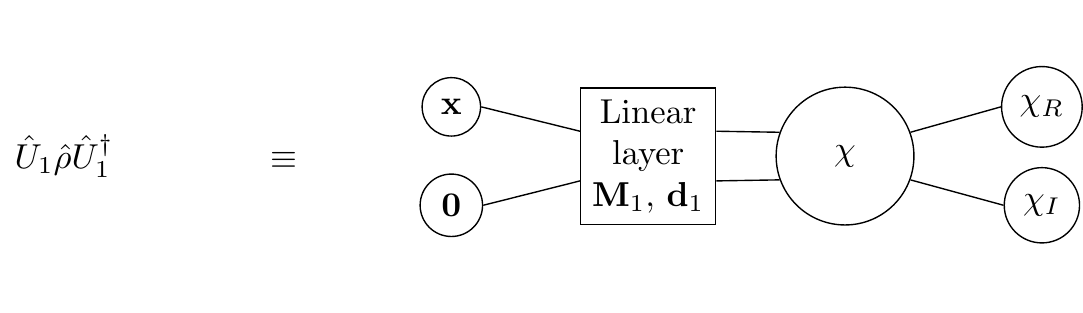}  \\
\includegraphics[width=0.8\textwidth]{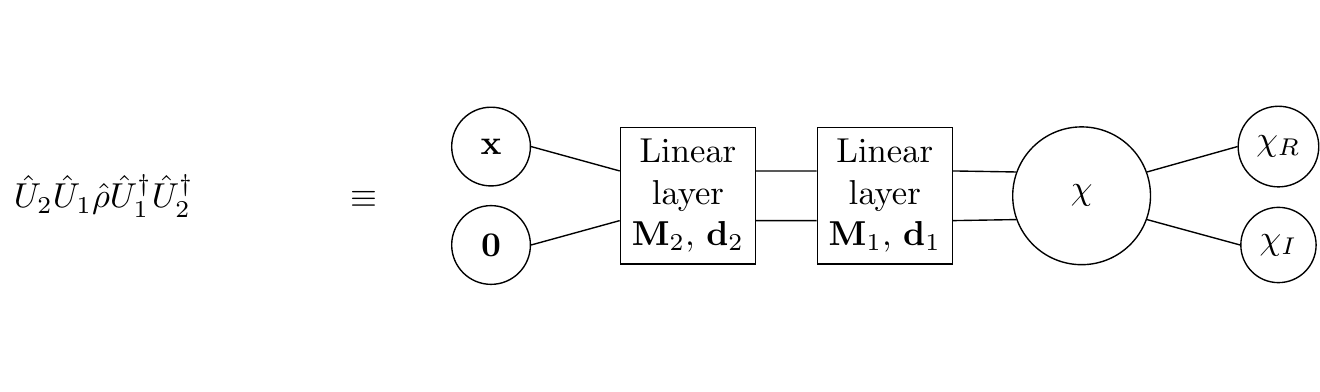}
\caption{Linear transformations and pullbacks.
  A single transformation with unitary operator $\hat{U}_1$ corresponds to a single pullback.
  A double transformation with unitary operators $\hat{U}_1$ and $\hat{U}_2$ corresponds to a double pullback.  Note that the flow of data from $\mbfx$ to the output through the network is in reverse order with respect to the two transformations.\label{fig:chpPBK9}}
  \end{figure*}

\begin{figure*}
 \centering
  \includegraphics[width=0.8\textwidth]{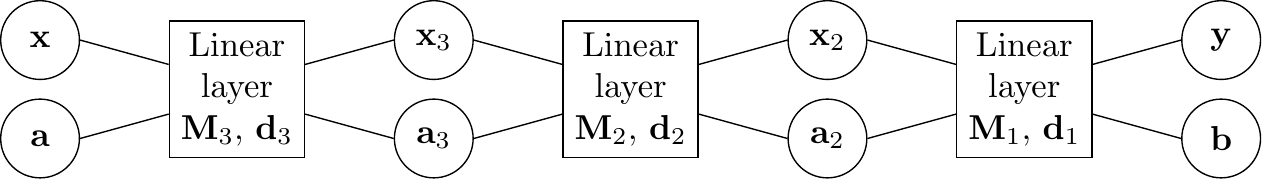}
  \caption{Example of cascading three transformations. Note the reverse order of layers from right to left.\label{fig:pllbck3}}
  \end{figure*}

The pullback is helpful when being in the presence of multiple transformations.
A sequence of linear transforms is equivalent to a sequence of pullbacks in reverse order as sketched in figure~\ref{fig:chpPBK9}.
%%%%%%%%%%%%%%%%%%%%%%%%%%%%%%%%%%%%%%%%%%%%%
%%%%%%%%%%%%%%%%%%%%%%%%%%%%%%%%%%%%%%%%%%%%%
For example, we consider a system originally described by the density matrix $\rho$ and canonical observables $\hat{\mbfR}$. First, the system is subject to a linear transformation with operator $\hat{U}_1$, such that the density matrix becomes
  \begin{equation}
    \label{eq:6}
    \hat{U}_1\rho\hat{U}^\dagger_1
  \end{equation}
and the new canonical vector is
\begin{equation}
  \label{eq:7}
  \hat{R}_1=\mbfM_1\hat{R}+\mbfd_1.
\end{equation}
We then consider a second transformation with unitary operator $\hat{U}_2$ and parameters $\mbfM_2$ and $\mbfd_2$, such that the final
density matrix reads
\begin{equation}
  \label{eq:8}
  \hat{U}_2\hat{U}_1\rho\hat{U}_1^\dagger\hat{U}_2^\dagger  ,
\end{equation}
and the final vector of observables is
\begin{equation}
  \label{eq:9}
  \hat{\mbfR}_2=\mbfM_2 \hat{\mbfR}_1+\mbfd_2=
  \mbfM_2\mbfM_1 \hat{\mbfR}+\mbfd_2+\mbfM_2\mbfd_1\text{.}
\end{equation}
We remark that $\mbfM_1\mbfM_2\neq\mbfM_2\mbfM_1$, hence the
order of the two tranformations is relevant as the
corresponding unitary operators $\hat{U}_1$ and $\hat{U}_2$ do not commute.
The sequence of the two linear transformations is equivalent to a single transformation with parameters
\begin{equation}
  \begin{array}{l}
    \label{eq:10}
    \mbfM=\mbfM_2 \mbfM_1\\
    \mbfd'=\mbfM_2\mbfd_1+\mbfd_2
    \end{array}
  \end{equation}
%%%%%%%%%%%%%%%%%%%%%%%%%%%%%%%%%%%%%%%%%%%%%
\begin{figure*}
  \centering
  \includegraphics[width=0.8\textwidth]{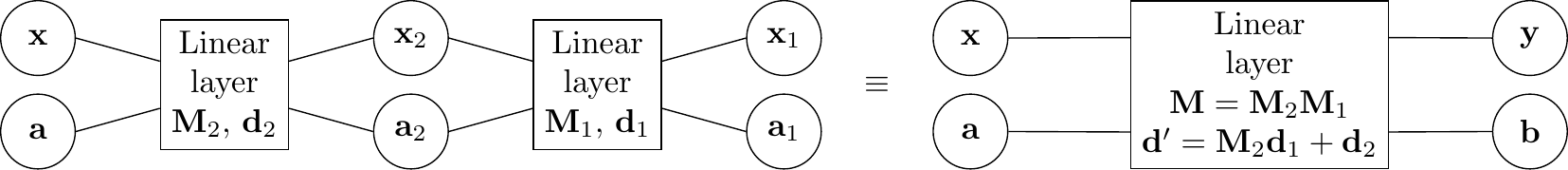}
  \caption{Equivalence of two cascaded linear layers with
    a single linear layer.\label{fig:chpPBKfig10}}
  \end{figure*}
%%%%%%%%%%%%%%%%%%%%%%%%%%%%%%%%%%%%%%%%%%%%%
We have the following (figure~\ref{fig:chpPBKfig10})
{\it Proposition ---    The sequence of the two linear transfomations $\hat{U}_1$ and $\hat{U_2}$ is equivalent to the pullback of the two linear layers with parameters $(\mbfM_1, \mbfd_1)$ and  $(\mbfM_2, \mbfd_2)$.}
\begin{proof}
As indicated in Fig.~\ref{fig:chpPBKfig10} the ouput of the linear layer corresponding to $\hat{U}_2$ is
    \begin{equation}
      \begin{array}{l}
        \mbfx_2=\mbfx \mbfM_2\\
        \mbfa_2=\mbfM_2^{-1}\left(\mbfd_2+\mbfa\right)
      \end{array}
    \end{equation}
    The output of the linear layer corresponding to $\hat{U}_1$ is
    \begin{equation}
      \begin{array}{l}
        \mbfx_1=\mbfx_2\mbfM_1=\mbfx \mbfM_2\mbfM_1\\
        \mbfa_1=\mbfM_1^{-1}\left(\mbfd_1+\mbfa_2\right)=
        \mbfM_1^{-1}\left(\mbfd_1+\mbfM_2^{-1}\mbfd_2+\mbfM_2^{-1}\mbfa\right)
        \label{eq:th2proof1}
      \end{array}
    \end{equation}
    For the layer with parameters $(\mbfM,\mbfd')$ we have, see~\eqref{eq:10},
    \begin{equation}
      \begin{aligned}
        \mbfy&=\mbfx \mbfM=\mbfx \mbfM_2 \mbfM_1\\
        \mbfb&=\mbfM^{-1}\left(\mbfd'+\mbfa\right)\\
        &=\mbfM_1^{-1}\mbfM_2^{-1}\left(\mbfM_2\mbfd_1+\mbfd_2+\mbfa\right)\\
        &=\mbfM_1^{-1}\left(\mbfd_1+\mbfM_2^{-1}\mbfd_2+\mbfM_2^{-1}\mbfa\right)
      \end{aligned}
      \label{eq:th2proof2}
    \end{equation}

As $\mbfy=\mbfx_1$ and $\mbfb=\mbfa_1$ we have the proof.
    \end{proof}
By extending the previous argument to three or more transformations, one realizes that the cascade of an arbitrary number of transformations corresponds to a cascade of pullbacks in reverse order.  For $M$ transformations, reverse order means that one first make a pullback of operator $1$,  then operator $2$, etc. In the flow of data in the network, the input $\mbfx$ first enters operator $M$, then $M-1$ and so forth to passing through the linear layer corresponding to the transformation $1$.  The case $M=3$ is shown as an example in figure~\ref{fig:pllbck3}.
%%%%%%%%%%%%%%%%%%%%%%%%%%%%%%%%%%%%%%%%%%%%%
%%%%%%%%%%%%%%%%%%%%%%%%%%%%%%%%%%%%%%%%%%%%%
\section{The Glauber displacement layer}\label{sec:glaub-displ-layer} Starting from the linear layer, we can define specialized layers corresponding to
  different unitary operators. The first we describe is the displacement operator,
  or Glauber operator, defined by~\cite{BarnettBook}
\begin{equation}
\label{eq:11}
\hat{\mathcal{D}}(\alpha)=\exp\left(\alpha^*\hat{a}-\alpha\hat{a}^\dagger\right)\;.
\end{equation}
A coherent state is obtained by displacing the vacuum state:
\begin{equation}
   \label{eq:12}
  |\alpha\rangle=\hat{\mathcal{D}}(\alpha)|0\rangle.
\end{equation}
For a single mode, letting $\hat{U}=\hat{\mathcal{D}}(\alpha)$, we have
\begin{equation}
\hat{\widetilde{a}}=\hat{U}^\dagger\hat{a}\hat{U}=\hat{a}+\alpha,
\end{equation}
 which implies for the canonical vector
 \begin{equation}
 \hat{\widetilde{\mbfR}}= \hat{U}^\dagger\hat{\mbfR}\hat{U}=
 \begin{pmatrix}\hat{x}\\\hat{p}\end{pmatrix}+
     \begin{pmatrix}d_0\\d_1\end{pmatrix}
   \end{equation}
with
\begin{equation}
  \label{eq:14}
  \begin{array}{l}
    d_0=\sqrt{2}\Re(\alpha)\\
    d_1=\sqrt{2}\Im(\alpha).
    \end{array}
\end{equation}
For a many-body displacement $\hat{\mathcal{D}}(\bm{\alpha})$,
      wit ${\bm{\alpha}}=(\alpha_0,\alpha_1,...,\alpha_n)^\top$,
      we have
  \begin{equation}
    \hat{\widetilde{\mbfa}}=\hat{U}^\dagger\hat{\mbfa}\hat{U}=\hat{\mbfa}+
    \bm{\alpha},
    \end{equation}
    which implies for the canonical vector
    \begin{equation}
   \hat{\widetilde{\mbfR}}=\hat{U}^\dagger\hat{\mbfR}\hat{U}=
   \hat{\mbfR}+\mbfd
      \end{equation}
      with ($j=0,1,\ldots,n-1$)
\begin{equation}
  \label{eq:13}
  \begin{aligned}
      d_{2j}&=\sqrt{2}\Re(\alpha_j)\\
      d_{2j+1}&=\sqrt{2}\Im(\alpha_j)
    \end{aligned}
  \end{equation}
 For the Glauber layer we hence have $\mbfM={\bf 1}_N$.

\begin{figure}
  \centering
  \includegraphics[width=0.45\textwidth]{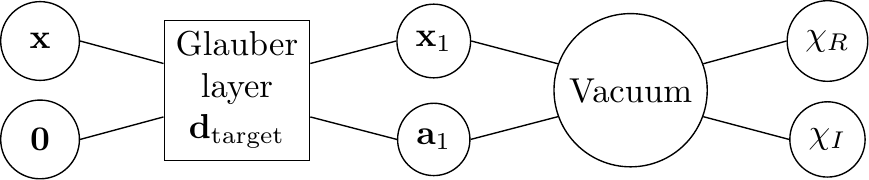}
  \caption{Model for a coherent state obtained by pullback of a displacement operator (Glauber layer) from the vacuum.\label{fig:chpPBKfig12}}
  \end{figure}

 To create a NN model that represents a coherent state with
a given displacement vector $\mbfd_{\text{target}}$, one can start from the vacuum Gaussian state with $\mbfg={\bf 1}_N$ and $\mbfd={\bf 0}$ and pullback a linear gate with $\mbfM={\bf 1}_N$, and displacement $\mbfd_{\text{target}}$. No bias is needed (see Fig.~\ref{fig:chpPBKfig12} with $\mbfa={\bf 0}$).
%%%%%%%%%%%%%%%%%%%%%%%%%%%%%%%%%%%%%%%%%%%%%
\section{The squeezing layer}
Different nonclassical states are generated from vacuum by unitary operators resulting into the following linear relation
\begin{equation}
  \hat{\widetilde{\mbfa}}=\mbfU \hat{\mbfa}+\mbfW \hat{\mbfa}^\dagger,
  \label{eq:gensym}
\end{equation}
which generalizes Eq.~({\ref{eq:lineartransform}}).
Using the matrices $\mbfU$ and $\mbfW$, one obtains the corresponding symplectic matrix $\mbfM$.

\noindent Let
\begin{equation}
\mbfU=\mbfU_R+\imath\mbfU_I
\end{equation}
and
\begin{equation}
\mbfW=\mbfW_R+\imath\mbfW_I.
\end{equation}
We write the $\hat{\widetilde{\mbfa}}$ as follows

\begin{eqnarray}
\hat{\widetilde\mbfa} &=&\displaystyle \frac{\hat{\widetilde\mbfq}+\imath \hat{\widetilde\mbfp}}{\sqrt{2}}=\displaystyle\mbfU \hat{\mbfa}+\mbfW \hat{\mbfa}^\dagger\\
\hat{\widetilde\mbfa}^\dagger &=&\displaystyle \frac{\hat{\widetilde\mbfq}-\imath \hat{\widetilde\mbfp}}{\sqrt{2}}=\mbfU^* \hat{\mbfa}^\dagger+\mbfW^* \hat{\mbfa}.
\end{eqnarray}
We have
\begin{equation}
\begin{array}{l}
    \displaystyle\hat{\widetilde\mbfq} =
  (\mbfU_R+\mbfW_R) \hat{\mbfq}+(-\mbfU_I+\mbfW_I) \hat{\mbfp}  \vspace{0.2cm}
\\
    \displaystyle\hat{\widetilde\mbfp} =
  (\mbfU_I+\mbfW_I) \hat{\mbfq}+(\mbfU_R-\mbfW_R) \hat{\mbfp}\vspace{0.2cm}
\end{array}
\end{equation}
which generalize Eq.~(\ref{eq:qptransform})
Following the same arguments for Eq.~(\ref{eq:symplecticmain}), we have for the symplectic matrix
\begin{equation}
  \begin{array}{l}
    \mbfM=\mbfM_1+\mbfM_2\vspace{0.2cm}\\
    \mbfM_1 = \mbfRq (\mbfU_R+\mbfW_R)\mbfRq^\top+\mbfRp (\mbfU_R-\mbfW_R)\mbfRp^\top\vspace{0.2cm}\\
    \mbfM_2 = \mbfRp (\mbfU_I+\mbfW_I)\mbfRq^\top+\mbfRq (-\mbfU_I+\mbfW_I)\mbfRp^\top\vspace{0.2cm}
\end{array}\label{eq:symplecticgeneral}
\end{equation}
Equation~(\ref{eq:symplecticgeneral}) is used for programming layers representing squeezing operators.
\subsection{Single-mode squeezed state}
We first consider a single model squeezed state, so that $N=2$, and
\begin{equation}
  \hat{\mbfR}=\begin{pmatrix}\hat{x}\\\hat{p}\end{pmatrix}
\end{equation}
Using the squeezing operator with parameters $r$ and $\theta$~\cite{BarnettBook}
\begin{equation}
  \label{eq:squeezeoperatorsinglemode}
\hat{\widetilde{a}}=\hat{S}^\dagger\hat{a}\hat{S}=\cosh(r)\hat{a}-e^{\imath \theta}\sinh(r)\hat{a}^\dagger
\end{equation}
we have that the matrix $\mbfU_{R,I}$ and $\mbfW_{R,I}$ are complex scalars as follows
\begin{equation}
  \label{}
  \begin{array}{l}
 U_R=\cosh(r)\\
 U_I=0\\
    W_R=-\cos(\theta)\sinh(r)\\
 W_I=-\sin(\theta)\sinh(r)
  \end{array}
\end{equation}
and, by \eqref{eq:symplecticgeneral}, we find the symplectic operator for the squeezing
%\footnote{See the $\matlab$ symbolic file {\tt matlabsymbolic/squeezedoperator.m}.}
\begin{equation}
  \label{eq:Msqueezingonemode}
  \begin{array}{ll}
   &\mbfM_s(r,\theta)=
                       \\[2pt]&\left(
  \begin{array}{cc}
         \cosh(r)-\cos(\theta)\sinh(r) & -\sin(\theta)\sinh(r) \\
            -\sin(\theta)\sinh(r) & \cosh(r)+\cos(\theta)\sinh(r)
  \end{array}\right)
                                    \end{array}
\end{equation}
For $\theta=0$, i.e., for a real squeezing parameter, we have
\begin{equation}
  \label{eq:Msqueezingonemode0}
\mbfM_s(r,0)=\begin{pmatrix} \exp(-r) & 0 \\ 0 & \exp(r)
\end{pmatrix}.
\end{equation}
\subsection{Multi-mode squeezed vacuum model}
\begin{figure}
  \centering%
  \includegraphics[width=0.4\textwidth]{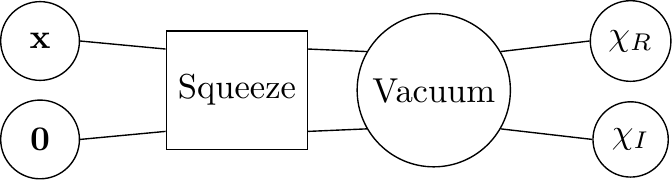}%
    \caption{A squeezed coherent state obtained by pullback with a single mode squeezing layer from the vacuum.\label{fig:chpSQUfig1}}
\end{figure}

We are interested to multi-mode systems, so we consider a $n-$body state, and apply the squeezing operator to one mode.
% \footnote{The {\tt Python} notebook for this section is  {\tt jupyternotebooks/phasespace/SingleModeSqueezer.ipynb}}.
%The $\hat{\mbfR}$ has dimension $N=2n$.
The single-mode squeezing operator is obtained by a linear gate with $\mbfd'=0$ and $\mbfM$ given by Eq.~(\ref{eq:Msqueezingonemode}) for the mode to be squeezed denoted by an index $n_\text{squeezed}$ in the range $0$ to $n-1$ (see Fig.~\ref{fig:chpSQUfig1}). For example, for $n=4$, and $n_\text{squeezed}=0$, we have
\begin{equation}
  \label{}
  \mbfM=\begin{pmatrix}
    M_{s,11} & M_{s,12} & 0 & 0 & 0 & 0 & 0 & 0 \\
    M_{s,21} & M_{s,22} & 0 & 0 & 0 & 0 & 0 & 0 \\
    0 & 0 & 1 & 0 & 0 & 0 & 0 & 0 \\
    0 & 0 & 0 & 1 & 0 & 0 & 0 & 0 \\
    0 & 0 & 0 & 0 & 1 & 0 & 0 & 0 \\
    0 & 0 & 0 & 0 & 0 & 1 & 0 & 0 \\
    0 & 0 & 0 & 0 & 0 & 0 & 1 & 0 \\
    0 & 0 & 0 & 0 & 0 & 0 & 0 & 1
    \end{pmatrix}.
\end{equation}
Seemingly for $n=4$, $n_\text{squeezed}=2$
\begin{equation}
  \label{}
  \mbfM=\begin{pmatrix}
    1 & 0 & 0 & 0 & 0 & 0 & 0 & 0 \\
    0 & 1 & 0 & 0 & 0 & 0 & 0 & 0 \\
    0 & 0 & 1 & 0 & 0 & 0 & 0 & 0 \\
    0 & 0 & 0 & 1 & 0 & 0 & 0 & 0 \\
    0 & 0 & 0 & 0 & M_{s,11} & M_{s,12} & 0 & 0 \\
    0 & 0 & 0 & 0 & M_{s,21} & M_{s,22} & 0 & 0 \\
    0 & 0 & 0 & 0 & 0 & 0 & 1 & 0 \\
    0 & 0 & 0 & 0 & 0 & 0 & 0 & 1
    \end{pmatrix}.
  \end{equation}
 %%%%%%%%%%%%%%%%%%%%%%%%%%%%%%%%%%%%%%%%%%%%%%%%%%%%%%%%%
 %%%%%%%%%%%%%%%%%%%%%%%%%%%%%%%%%%%%%%%%%%%%%%%%%%%%%%%%%
\subsection{Squeezed coherent states}
\begin{figure}
   \centering%
  \includegraphics[width=\columnwidth]{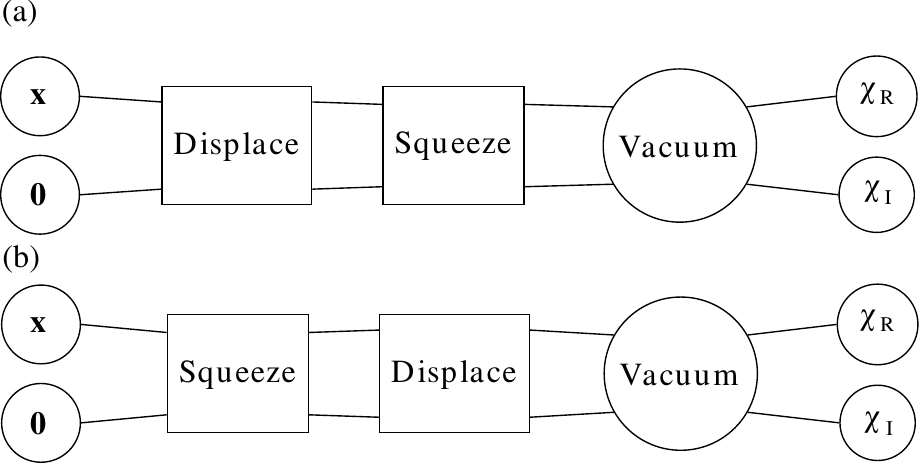}\\%
    \caption{Pullback model for a squeezed coherent state by displacing
      a squeezed vacuum (top panel). Pullback model for squeezed coherent state by squeezing a displaced vacuum (bottom panel). Here all the layers act on the same mode.
      \label{fig:chpSQUfig2}}
\end{figure}

The squeezed operator is represented by a linear gate,
we can cascade with other layers, such as the Glauber displacement layer.
Figure~\ref{fig:chpSQUfig2}a shows a model
to generate a squeezed coherent state from the squeezed vacuum
by a displacement layer.

The squeezed coherent states $|\alpha,\zeta\rangle$ are built by applying the displacement operator $\hat{\mathcal{D}}(\alpha)$ to a squeezed vacuum $\hat{\mathcal{S}}(\zeta)|0\rangle$, where $\zeta=r\,e^{\imath \theta}$ is the complex squeezing parameter, that is
\begin{equation}
  |\alpha,\zeta\rangle=\hat{\mathcal{D}}(\alpha)\hat{\mathcal{S}}(\zeta)|0\rangle
  \end{equation}

The resulting state has $\langle \hat{a} \rangle=\alpha$, and it is squeezed.
%%%%%%%%%%%%%%%%%%%%%%%%%%%%%%%%%%%%%%%%%%%%%%%%%%%%%%%%%%%%%
%%%%%%%%%%%%%%%%%%%%%%%%%%%%%%%%%%%%%%%%%%%%%%%%%%%%%%%%%%%%%%%%%%%%%
\subsection{Squeezing the displaced vacuum}
A different squeezed coherent state is obtained by squeezing a coherent state.
We first pullback a displacement layer from the vacuum to create the
coherent state,
and then pullback a squeezing layer, as shown in Fig.~\ref{fig:chpSQUfig2}b. This corresponds to the following equation~\eqref{eq:17}
\begin{equation}
  \label{eq:17}
|\alpha\cosh(r)-\alpha^*e^{\imath\theta}\sinh(r),\zeta\rangle=\hat{\mathcal{S}}(\zeta)\hat{\mathcal{D}}(\alpha)|0\rangle
  \end{equation}
  \noindent The result is a squeezed coherent state with the same eigenvalues for the covariance matrix as above, but the mean value of the displacement is changed~\cite{BarnettBook}, i.e.,
\begin{equation}
  \label{}
  \langle \hat{a}\rangle=\alpha\cosh(r)-\alpha^*e^{\imath\theta}\sinh(r).
\end{equation}
\section{Computing the Hamiltonian}
In this manuscript, we are interested in the state of a many-body Hamiltonians, which
correspond to classical solitons. We consider the Hamiltonian
\begin{equation}
  \hat{H}=-\sum_j \left[
\hpsid_j \hpsi_{j+1}+\hpsid_j \hpsi_{j-1}
\right]
+\frac{\gamma}{2}\hpsid_j\hpsid_j\hpsi_j\hpsi_j=\hat{K}+\hat{V}\;,
  \label{eq:qs:H}
\end{equation}
with the potential energy
\begin{equation}
\hat{V}=\frac{\gamma}{2}\sum_j \left(\hat{n}_j^2-\hat{n}_j\right)\;
\end{equation}
being $\hat{n}_i=\hpsid_i\hpsi_i$, and the kinetic energy
\begin{equation}
 \hat{K}=\sum_{ij}\omega_{ij}\hpsid_i\hpsi_j\;,
\end{equation}
being
$\omega_{ij}=-\delta_{i,j-1}-\delta_{i,j+1}$
with $i,j$ in $[0,n-1]$ and with homogeneous boundary conditions, $ \omega_{0,-1}=\omega_{n-1,n}=0$.

Starting from the vacuum state, one can build the NN model of an arbitrary state by multiple pullbacks. We compute the mean value of observable quantities as $\hat{H}$ and $\hat{\mathcal{N}}$ as derivatives of $\chi$ at $\mbfx=0$.
We have in symmetric ordering~\cite{BarnettBook}
\begin{equation}
  \langle \hat{K}\rangle =\left.\sum_{jk}\omega_{jk}
    \left[\frac{\partial^2}{\partial\alpha_j \partial(-\alpha_k^*)}\chi(\mbfalpha)-\frac{\delta_{jk}}{2}\chi\right]\right|_{\mbfalpha=0}
  \label{eq:Kchi}
  \end{equation}
  where $\mbfalpha=\left(\alpha_0\ldots\alpha_{n-1},\alpha^*_0\ldots\alpha^*_{n-1}\right)$,
  and $\sqrt{2}\alpha_j=x_{2j}+\imath x_{2j+1}$,
and for the interaction term
  \begin{equation}
    \langle \hat{V}\rangle=
    \frac{\gamma}{2}
    \sum_j \left.\frac{\partial^4\chi}{\partial\alpha_j^2 {\partial(-\alpha_j^*)}^2}
    -2\frac{\partial^2\chi}{\partial\alpha_j {\partial(-\alpha_j^*)}}
    +\frac{\chi}{2}\right|_{\mbfalpha=0}\text{.}
    \label{eq:nj2}
  \end{equation}
These quantities can be computed by using automatic differentiation on the NN model.

Specifically,
\begin{equation}
  \begin{aligned}
   \langle \hat{K} \rangle&=
   -\frac{1}{2}
                            \sum_{mn} \\
    &\left(\frac{\partial^2 \chi}{\partial q_m\partial q_n}+\frac{\partial^2 \chi}{\partial p_m\partial p_n}+ \chi\right)\omega_{mn}^R\\
   +&\left.\left(
        \frac{\partial^2 \chi}{\partial q_m\partial p_n}-\frac{\partial^2 \chi}{\partial p_n\partial p_q}\right)\omega_{mn}^I
    \right|_{\mbfx=0}
    \end{aligned}
 \end{equation}
where
\begin{equation}
  \label{}
\omega_{mn}=\omega_{mn}^R+\imath \omega_{mn}^I\;,
\end{equation}
$q_j=x_{2j}$ and $p_j=x_{2j+1}$ with $j=0,1,...,{N/2}-1$.

For the potential energy,
\begin{equation}
  \label{}
\langle \hat{V}\rangle =\sum_{nm}V_{nm}\langle \hat{a}_n^\dagger\hat{a}_m^\dagger \hat{a}_n \hat{a}_m\rangle\;,
\end{equation}
being, for a local interaction,
\begin{equation}
  \label{}
V_{nm}=\frac{1}{2}\delta_{nm}\;.
\end{equation}

For a mode  with index $j=0,1,\ldots,n-1$, the normal ordering product is
  \begin{equation}
    \langle \hat{n}_j^2\rangle=\langle \hat{a}^\dagger_j  \hat{a}_j  \hat{a}^\dagger_j \hat{a}_j\rangle=
    \langle \hat{a}^\dagger_j \hat{a}^\dagger_j \hat{a}_j   \hat{a}_j\rangle+
    \langle \hat{a}^\dagger_j  \hat{a}_j\rangle\;.
    \label{eq:nj2}
  \end{equation}
where we used $[\hat{a}_j,\hat{a}_j^\dagger]=\hat{a}_j \hat{a}^\dagger_j -\hat{a}^\dagger_j \hat{a}_j =1$.

We have
\begin{equation}
  \begin{aligned}
    \langle \hat{a}^\dagger_j  \hat{a}_j\rangle =\langle \hat{n}_j\rangle=
    \left.-\frac{\partial^2 \chi}{\partial z_j~\partial z_j^*}\right|_{\mbfz=0}   =\\
    -\left.\frac{1}{2}\left(\partial^2_{q_j} +\partial^2_{p_j}\right) \chi_R\right|_{\mbfx=0}-\frac{1}{2}\;.
    \end{aligned}
\end{equation}
and
\begin{equation}
  \begin{aligned}
  \langle \hat{a}^\dagger_j \hat{a}^\dagger_j \hat{a}_j   \hat{a}_j\rangle=
\left.  {\left(\frac{\partial}{\partial z_j}\right)}^2
  {\left(-\frac{\partial}{\partial z_j^*}\right)}^2\chi\right|_{\mbfx=0} \\
-2
\left.  \left(\frac{\partial}{\partial z_j}\right)
  \left(-\frac{\partial}{\partial z_j^*}\right)\chi\right|_{\mbfx=0}
+\frac{1}{2}\;.
  \end{aligned}
  \label{eq:4der}
  \end{equation}

Eq.~(\ref{eq:4der}) can be expressed in terms of the derivatives w.r.t. $q_j$ and $p_j$ as follows
\begin{equation}
  \begin{aligned}
  \langle \hat{a}^\dagger_j \hat{a}^\dagger_j \hat{a}_j   \hat{a}_j\rangle=
\left.\frac{1}{4}{\left(\partial_{q_j}^2+\partial_{p_j}^2\right)}^2\chi_R\right|_{\mbfx=0}\\
+\left.  {\left(\partial_{q_j}^2+\partial_{p_j}^2\right)}\chi_R\right|_{\mbfx=0}
    +\frac{1}{2}
    \end{aligned}
  \label{eq:4dere}
  \end{equation}
  where, as above, $q_j=x_{2j}$ and $p_j=x_{2j+1}$ with $j=0,1,...,{N/2}-1$.

The real part $\chi_R$ enters in Eq.~(\ref{eq:4dere}) as $\langle \hat{n}_j^2\rangle$ in Eq.~(\ref{eq:nj2})  is real valued.

In general, we need to evaluate the fourth order derivatives $\frac{\partial^4\chi}{\partial x_s\partial x_p\partial x_q\partial x_r}$,
  at $\mbfx=0$.  The computation of the fourth-order derivatives may be demanding as it grows with $N^4$. For the specific case of Gaussian states, it can be simplified as the fourth order derivatives can be expressed in terms of the covariance matrix $\mbfg$.

% Gaussian states are determined by the covariance matrice $\mbfg$ and the displacement $\mbfd$.
We have by direct derivation of \eqref{eq:chiGaussian1}
\begin{equation}
  \label{eq:d4gaussian}
  \begin{aligned}
  \left.\frac{\partial^4 \chi}{\partial x_s \partial x_p \partial x_q \partial x_r}\right|_{\mbfx=0}\\
  &=\frac{1}{4}g_{sp}g_{qr}+
  \frac{1}{4}g_{sq}g_{pr}+
  \frac{1}{4}g_{sr}g_{pq}\\
  &+\frac{1}{2}g_{sp}d_q\,d_r
  +\frac{1}{2}g_{sq}d_p\,d_r
  +\frac{1}{2}g_{sr}d_p\,d_q\\
  &+\frac{1}{2}g_{pq}d_r\,d_s
  +\frac{1}{2}g_{pr}d_q\,d_s
  +\frac{1}{2}g_{qr}d_p\,d_s\\
  &+d_s d_p d_q d_r\;.
  \end{aligned}
  \end{equation}
When considering the diagonal terms we have
  (no implicit sum is present in the following equation)
\begin{equation}
  \left.\frac{\partial^4 \chi}{\partial x_q^4}\right|_{\mbfx=0}=
  \frac{3}{4}g_{qq}^2
  +3\,g_{qq}d_q^2+d_q^4\;.
  \label{eq:d4gaussiandiag}
  \end{equation}
Seemingly, we have
\begin{equation}
  \begin{aligned}
  \left.\frac{\partial^4 \chi}{\partial x_q^2\partial x_p^2}\right|_{\mbfx=0}=
  \frac{1}{4}g_{qq}g_{pp}+  \frac{1}{2}g_{pq}^2+\\
  \frac{1}{2}g_{pp}d_q^2
  +\frac{1}{2}g_{qq}d_p^2
  +2 g_{pq}d_p\,d_q+d_p^2 d_q^2\;.
  \end{aligned}
  \label{eq:d4gaussiandiag2}
  \end{equation}
Hence, for Gaussian states, we do not need to evaluate explicitly the fourth order derivatives, but we can combine the componentes of $\mbfg$ and $\mbfd$.
For the second derivatives we have
\begin{equation}
\begin{array}{l}
\displaystyle\left.\frac{\partial^2{\chi_R}}{{\partial x_m}{\partial x_n}}\right|
_{\mbfx=0}=
-\frac{1}{2}g_{mn}-d_m d_n\\
\displaystyle\left.\frac{\partial^2{\chi_I}}{{\partial x_m}{\partial x_n}}\right|
_{\mbfx=0}\textbf{}=
0\\
\end{array}\;.
\end{equation}
\section{Quantum soliton variational ansatz}
\begin{figure*}[ht!]
\includegraphics[width=\textwidth]{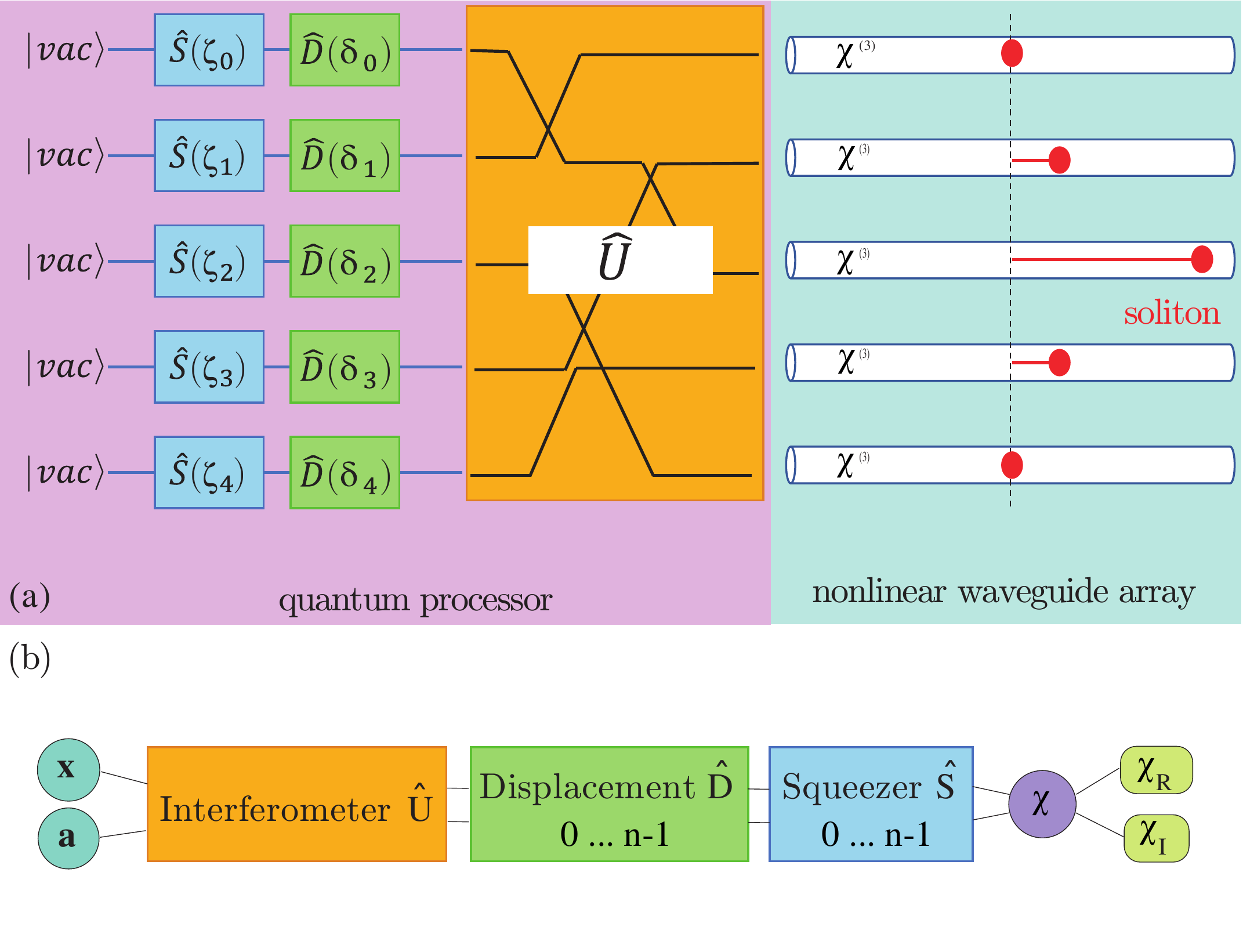}
  \caption{(a) Quantum processor for solitons in coupled optical waveguides. $n$ vacuum spatial modes are squeezed, displaced and mixed by a trainable interferometer $\hat{U}$ ($n=5$ in the figure).
    (b) Multiple pullbacks as a quantum soliton variational ansatz.~\label{fig:1}}
\end{figure*}

Figure~\ref{fig:1} shows the quantum processor to synthesize quantum lattice solitons in a discrete array with $n$ sites. The circuit is composed by $n$ squeezers with complex squeezing parameters $\zeta_j$, and displacement operators with complex displacements $\delta_j$, being $j=0,1,\ldots,n-1$ hereafter. A unitary interferometer $\hat{U}$ mixes the modes before entering the waveguide array. The quantum processor is represented as a NN in the phase space. All the parameters of the squeezers, displacements, and interferometers are trainable and initialized as random variables. We refer to the state generated by the quantum circuit as the quantum soliton variational ansatz (QSVA).
%%%%%%%%%%%%%%%%%%%%%%%%%%%%%%%%%%%%%%%%%%%%%
%%%%%%%%%%%%%%%%%%%%%%%%%%%%%%%%%%%%%%%%%%%%%
We build the NN model of an arbitrary state by multiple pullbacks from the vacuum.
Figure~\ref{fig:1}b shows the NN model for the QSVA in Fig.~\ref{fig:1}a.

We obtain the mean values of $\hat{H}$ and $\hat{\mathcal{N}}$ as derivatives
of $\chi$. As described above, for a Gaussian state, derivatives are obtained algebraically by $\mbfg$ and $\mbfd$, which speeds up training by conventional steepest descent. We use as cost function $\exp(\langle \hat{H}\rangle /n)$, ($\langle \hat{H}\rangle<0$ at a minimum).  As additional cost function, we use ${(\langle \mathcal{\hat{N}}\rangle -N_{T})}^2$, to constraint the target mean boson number as $\langle \mathcal{\hat{N}}\rangle=N_T$.

The first challenge is training the variational quantum circuit to generate single quantum solitons. The simplest soliton is localized in one of the sites (as $\gamma<0$).  We want to understand if the trained QSVA can approximate a quantum version of the single soliton, such that one can use quantum processors to generate quantum solitons and investigate their physics.

We find that the NN furnishes quantum solitons with lower energy than the classical solution so that the quantum solitons form a larger class of nonlinear waves.
Starting from randomly generated weights, the soliton is found after training the NN model which minimizes the cost function, and $\langle \hat{H}\rangle$ at fixed $\langle\hat{\mathcal{N}}\rangle=N_{T}$.

\begin{figure*}[ht!]
    \includegraphics[width=\textwidth]{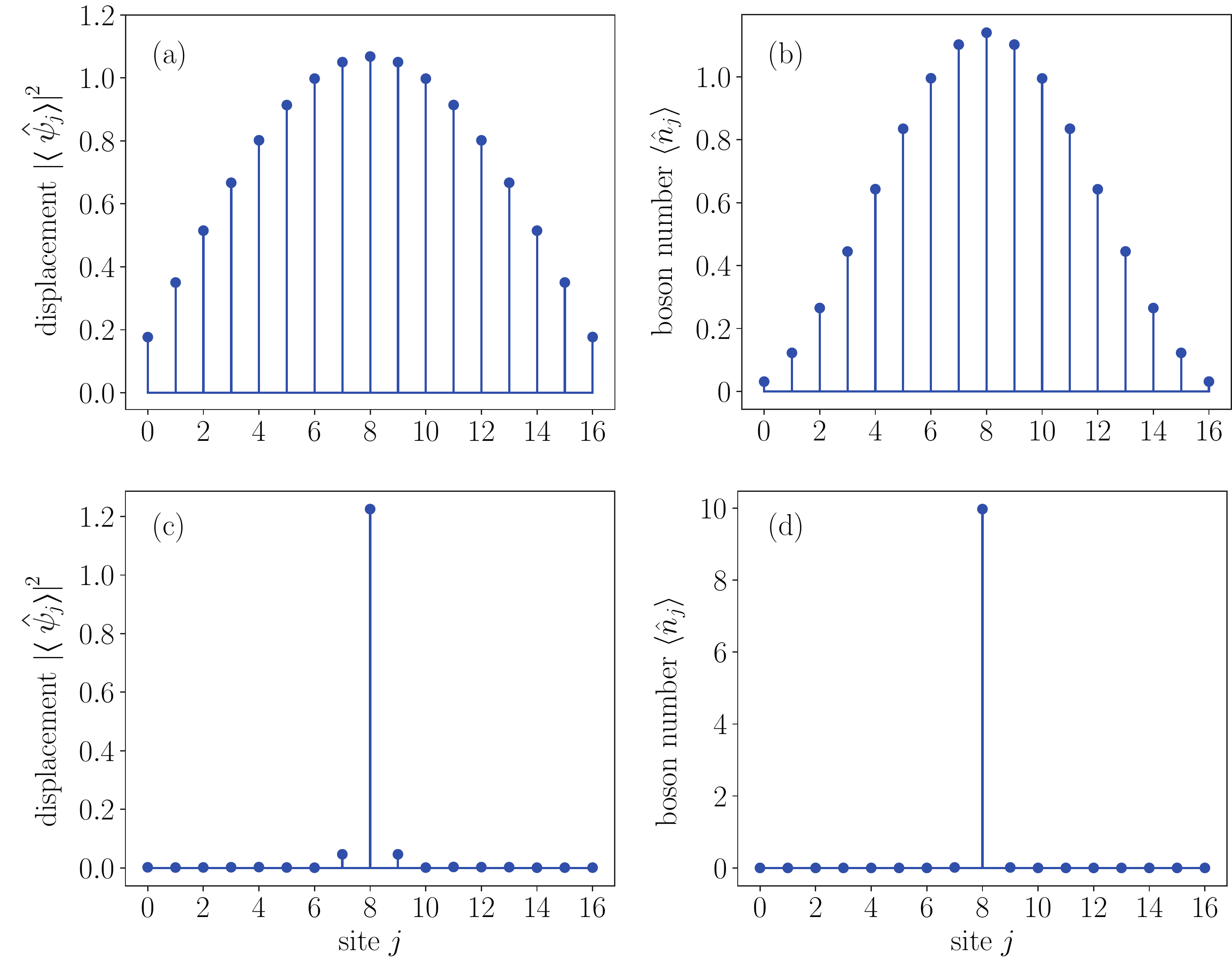}
    \caption{\label{fig:N10gamma-1} (a,b)  Ground state for $N_{T}=10$ and $\gamma=-0.01$: (a) displacement $|\langle \hat{\psi}_{j}\rangle|^{2}$; (b) mean boson number $\langle \hat{n}_{j} \rangle$; (c,d) as in (a,b) with $N_{T}=10$ and $\gamma=-1$ ($n=17$).}
  \end{figure*}

We first consider small interaction strength $|\gamma|$, at which we expected delocalized solutions. We show in Fig.~\ref{fig:N10gamma-1}a the displacements $|\langle\psi_j\rangle|^2=(d_{2j}^2+d_{2j+1}^2)/2$ after the training.
In Fig.~\ref{fig:N10gamma-1}b we show the mean boson number $\langle \hat{n}_j \rangle$, for $N_{T}=10$ and $\gamma=-0.01$ after thousands of epochs; the profile is delocalized.
Strong localization is obtained for $N_{T}=10$ and $\gamma=-1$ as in Fig.~\ref{fig:N10gamma-1}b,c.

%%%%%%%%%%%%%%%%%%%%%%%%%%%%%%%%%%% 
%%%%%%%%%%%%%%%%%%%%%%%%%%%%%%%%%%%%%%%
\begin{figure*}[ht!]
  \includegraphics[width=\textwidth]{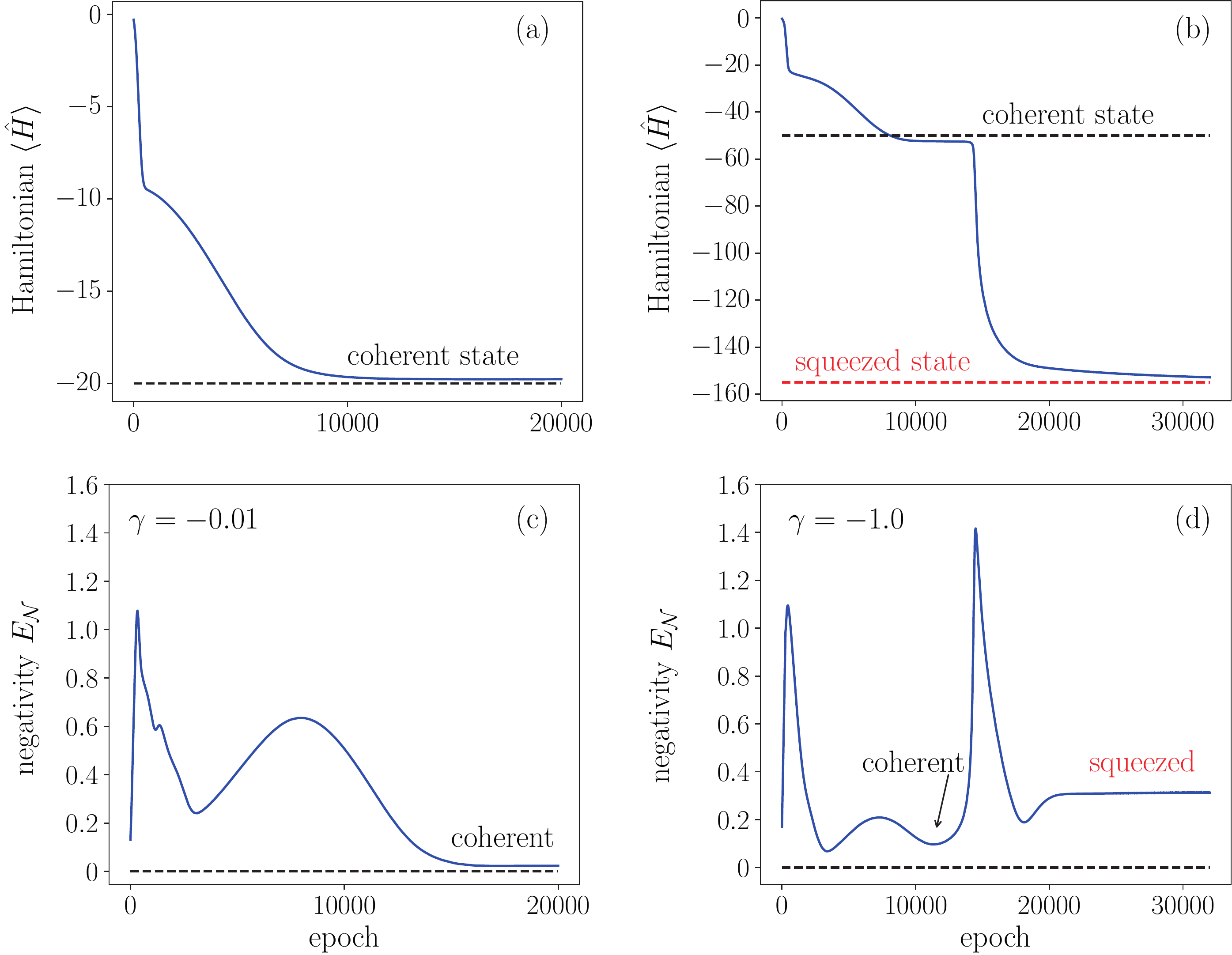}
\caption{\label{fig:NtrainingN10gamma-1}
  Training history for Fig.~\ref{fig:N10gamma-1}: (a) $\langle H\rangle$ rapidly reaches the minimum for $\gamma=-0.01$ (dashed line is the coherent state with $\langle \hat{H}\rangle\simeq -2N_{T}$); (b) as in (a) for $\gamma=-1$. Dashed line is Eq.~(\ref{eq:1}).  The model explores plateaus corresponding to solitons with different degree of entanglement;
  (c) logarithmic negativity for $\gamma=-0.01$, no entanglement (dashed line); (d) as in (c) for $\gamma=-1$, the ground state is entangled.}
  \end{figure*}
%%%%%%%%%%%%%%%%%%%%%%%%%%%%%%%%%%%%%
In figure~\ref{fig:NtrainingN10gamma-1}, we report $\langle\hat{H}\rangle$ when varying the number of training epochs for $N_{T}=10$. The algorithm converges after some thousands of epochs, and explores different solutions.
Figure~\ref{fig:NtrainingN10gamma-1}a shows the training when $\gamma=-0.01$: the NN converges to the delocalized state in Fig.~\ref{fig:N10gamma-1}a,b. To compare with analytical estimates, we consider the continuum limit and a non-squeezed coherent
state with a sinusoidal profile. The value of the Hamiltonian $\langle \hat{H}\rangle\simeq -2 N_T$ (dashed line in fig.~\ref{fig:NtrainingN10gamma-1}a).
Figure~\ref{fig:NtrainingN10gamma-1}b shows the strong interaction case $\gamma=-1$.
During the training the systems settles in two plateaus corresponding to $\langle \hat{H}\rangle\cong -50$  and $\langle \hat{H}\rangle\cong -155$.
These values are understood in terms of approximate solutions, in which
$\langle \hat{\psi}_{j\neq A}\rangle=0$, with $A=\left\lfloor\frac{n}{2}\right\rfloor+1$,
where the soliton is localized. If we consider the many-body state
$|\mbfalpha\rangle=\hat{D}_A(\mbfalpha)|\text{\bf vac}\rangle$  with $\alpha_i=\delta_{iA}\alpha$, we have
  \begin{equation}
    \langle \hat{H}\rangle\simeq \frac{\gamma}{2}\left(
      \sum_i \langle|\hat{n_i}^2|\rangle- \langle|\hat{n_i}|\rangle\right)
    =\frac{\gamma}{2}\sum_i |\alpha_i|^4=\frac{\gamma}{2}|\alpha|^4
    \end{equation}
One has $\langle \hat{N}\rangle = |\alpha|^2=N_T$ and $\langle \hat{H}\rangle=\frac{\gamma}{2} N_T^2$
which gives  $\langle \hat{H}\rangle\simeq -50$ (first plateau in Fig.~\ref{fig:NtrainingN10gamma-1}, dashed line).
The difference due to the neglected $\langle \hat{K} \rangle$. %$|\mbfalpha\rangle$ has a negligible entanglement.

The lowest energy solution occurs at a larger number of epochs (Fig.~\ref{fig:NtrainingN10gamma-1}b).
The plateau can be estimated by considering a squeezed coherent state
  $|\mbfalpha,\zeta\rangle=\hat{D}_A(\mbfalpha)\hat{S}_A(\zeta)|\textbf{vac}\rangle$
% \begin{equation}
%   |\mbfalpha,\zeta\rangle=\hat{D}_A(\mbfalpha)\hat{S}_A(\zeta)|\textbf{vac}\rangle
%   \end{equation}
  where $\hat{S}_A$ is the squeezing operator at $j=A$ with $\zeta=r e^{\imath\theta}$.
  For $|\mbfalpha,\zeta\rangle$ one has
  \begin{equation}
    \begin{aligned}
      \langle \mathcal{\hat{N}}\rangle&={\sinh(r)}^2+|\alpha|^2=N_T\\
      \langle \hat{H}\rangle&=\frac{5}{8}-2|\alpha^2|+|\alpha|^4+
      (-1+2|\alpha|^2)\cosh(2 r)+\\&\frac{3}{8}\cosh(4 r)-|\alpha|^2\cos(2\phi-\theta)\sinh(2r ),
    \end{aligned}
    \label{eq:1}
    \end{equation}
with minimum $\langle \hat{H}\rangle\simeq -155$ ($N_T=10$, red dashed line).
\section{Entangled solitons} Having evidence that the ground state is not the classical solution, we are interested to understand if other non-classical features are present. We consider precisely the degree of entanglement.

Entanglement in Gaussian states is estimated by the logarithmic negativity~\cite{Vidal2002, Illuminati2004,Plenio2007, Martynov2020}. We partition the system in the soliton site, ``Alice'' A at $A=\left\lfloor\frac{n}{2}\right\rfloor+1$, and the sites with $j\neq A$, ``Bob''.

We retrieve the covariance matrix $\mbfg$ from the trained NN model,
and its partial transpose  $\tilde\mbfg$~\cite{Vidal2002},
obtained by multiplying by $-1$ the elements of Alice momenta $\hat{p}_A$.
The symplectic eigenvalues $\tilde c_0,\tilde c_1,\ldots,\tilde c_{n-1}$
are moduli of the eigenvalues of ${\bf J}^\top\tilde\mbfg/2$, and
logarithmic negativity is
\begin{equation}
  E_{\mathcal{N}}=-\sum_{j=0}^{n-1}\log_2\min\{1, 2 c_j\}.
  \end{equation}

Figure~\ref{fig:NtrainingN10gamma-1}c shows the $E_\mathcal{N}$ for $\gamma=-0.01$ during the training. Despite
  initially the system explores randomly generated entangled solution, the asymptotic state is a coherent state, with vanishing entanglement (dashed line) as expected for a negligible interaction.
  
  Figure~\ref{fig:NtrainingN10gamma-1}d shows $E_\mathcal{N}$ for $\gamma=-1$.
  The local minimum of $\langle \hat{H}\rangle$, the coherent state, has a smaller entanglement than the squeezed ground state at epoch~$3\times 10^4$.
  Nevertheless, the asymptotic value of $E_\mathcal{N}$ is negligible if compared with the bounded solitons considered in the following.

  Given the non-classical features in the single-soliton ground state, we study the broader family of multiple solitons. We expect that strong nonlinear localization in different channels increases the degree of entanglement. We study two-soliton bound states~\cite{Lederer2008, Torner2011, KivsharBook}, which can be approximated by training the network by proper loss functions.
We fix the location of two sites, denoted $A$ and $B=A+\Delta$, and we introduce as cost function
$\exp(-\langle \hat{n}_{A}\rangle)$ to maximize the bosons at $A$ and $\exp(\langle \hat{n}_{A}-\hat{n}_{B}\rangle)$
to generate a solution with the same boson number in $B$.

\begin{figure*}[ht]
\includegraphics[width=\textwidth]{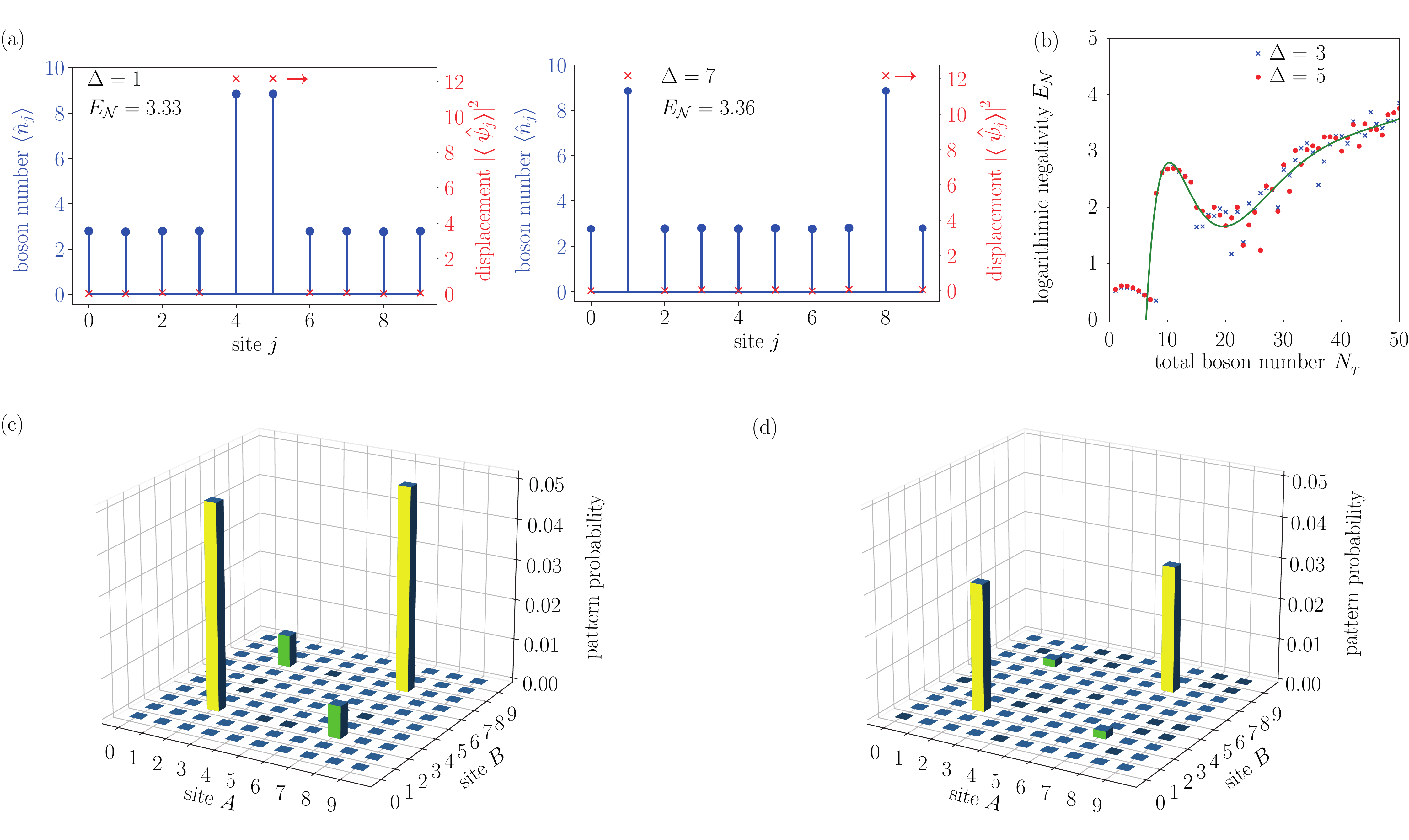}
\caption{\label{fig:TwinN40}   (a)  Boson distribution and mean displacement (right axis) for two representative bound-soliton solutions with $N_{T}=40$ and $\gamma=-1.0$; (b) logarithmic negativity versus the boson number $N_T$ for different $\Delta$;
  (c) probabilities for events containing $\bar{n}_T=2$ particles in two varying sites $A$ and $B$ and $N_T=10$; (d) as in (c) with 2 particles per site ($\bar{n}_T=4$). The probability is not vanishing only when the particles are at the soliton sites, denoting the generation of entangled pairs ($n=10$).}
  \end{figure*}

Figure~\ref{fig:TwinN40} shows the coupled solitons for different peak positions ($N_T=40$).
At variance with the single soliton in Fig.~\ref{fig:N10gamma-1}, the boson number is not vanishing in all the
sites, but ${|\langle \psi_j\rangle|}^2$ is not negligible only at the solitons, outlining the onset of squeezed nonlocal states.

The bound solitons are entangled. If we consider a bipartite system composed by the site in $A$ and the rest of the array, $E_{\mathcal{N}}$ depends on the distance $\Delta$ between the two solitons (Fig.~\ref{fig:TwinN40}).
The logarithmic negativity depends on the number of total bosons $N_T$:
Figure~\ref{fig:TwinN40}b shows $E_{\mathcal{N}}$ versus $N_T$ for various $\Delta$.
The entanglement is small and comparable with the single soliton for case $N_T< 10$, and growing when $N_T> 10$.
%%%%%%%%%%%%%%%%%%%%%%%%%%%%%%%%%%%%%%%
\section{Boson sampling} As we show in the following the entanglement is connected to correlated particles in different output channels. We compute the probability $\Pr(\bar\mbfn)$ of finding $\bar{n}_0$ bosons in mode $0$, $\bar{n}_1$ in mode $1$, and so forth, where $\bar\mbfn=(\bar{n}_0,\bar{n}_1,\ldots,\bar{n}_{n-1})$ is a given particle pattern. Letting $\rho$ the density matrix, one has $\Pr(\bar\mbfn)=\Tr[\rho|\bar\mbfn\rangle\langle\bar\mbfn|]$, with $|\bar\mbfn\rangle\langle\bar\mbfn|=\otimes_j|\bar{n}_j\rangle\langle \bar{n}_j|$.
Correspondingly~\cite{Kruse2018},
\begin{equation}
  \Pr(\bar\mbfn)=\left.\frac{1}{\bar{\mbfn}!}
\prod_j  {\left(\frac{\partial^2}{\partial \alpha_j\partial \alpha_j^*}\right)}^{\bar{n}_j}
    e^{\sum_j|\alpha|_j^2}Q_\rho(\mbfalpha)\right|_{\mbfalpha=0} 
\end{equation}
where $\bar\mbfn!=\bar{n_0}!\bar{n}_1!\ldots\bar{n}_{n-1}!$
and $Q_\rho=\pi^n \langle \mbfalpha | \rho | \mbfalpha \rangle$
is the Q-representation of the density matrix~\cite{GardinerBook, BarnettBook}.

We obtain $\Pr(\bar\mbfn)$ as in~\cite{ContiQMI2021}.
We consider an event with $\bar{n}_T$ particles in two sites $A$ and $B$.
Specifically, the events such that (i) $\bar{n}_A=\bar{n}_B=n_T/2$ for $A\neq B$, or (ii) $\bar{n}_A=\bar{n}_B=\bar{n}_T$ when $A=B$, and $\bar{n}_j=0$ elsewhere. For $n_T=2 $ this corresponds to observing a pair of particles in a single site, or two particles in two distinct sites.

Figure~\ref{fig:TwinN40}c shows the probability for $\bar{n}_T=2$ varying $A$ and $B$ when $N_T=10$.
Such a probability is different from zero only when $A$ or $B$
are at a soliton position ($A=2$ and $B=7$ in Fig.~\ref{fig:TwinN40}c).
When $A=B$ (yellow bar) we have the probability of observing the two particles in one single soliton.
When $A$ and $B$ corresponds to the sites of the two solitons (green bar),
we have the probability of observing a couple of particles in the
two soliton sites. For all other patterns with $\bar{n}_T=2$ there is
a negligible probability of observing a pair. Hence, pairs of particles appear either in a single soliton,
or entangled in the two solitons. When $\bar{n}_T=4$ we have non-vanishing probability of observing either $4$ bosons in a soliton, or two entangled pairs in the two solitons~(Fig.~\ref{fig:TwinN40}d).
Particles are hence observed simultaneously only when they are monitored at the two localized solitons, demonstrating quantum correlation and the nonlocality.
\section{Conclusion}\label{sec:conclusion}
Discrete waveguide arrays became popular for the quantum advantage in linear boson sampling.
One can argue about the role of nonlinearity, which -- classically-- supports discrete self-localized solitons.

For unveiling quantum effects in discrete solitons, we adopted quantum machine learning.
Variational quantum circuits enable the theoretical investigation of quantum nonlinear waves and their bound states. Also, their physical realization as tunable quantum processors opens to new experiments on non-perturbative nonlinear regimes.

We considered Gaussian states as they are expected to conform to experimental observations at high intensities, and also
allowed us to rigorously obtain the degree of entanglement. Furthermore, the neural network formulation is compatible with the Gaussian boson sampling protocol in the presence of nonlinearity.

We found that solitonic states alter the observation of multi-particle events.
Entangled pairs emerge from self-localized bounded solitons, which can be synthesized by trained  quantum circuits.

Quantum machine learning and quantum variational algorithms open new possibilities for the physics and the applications of quantum nonlinear waves.
The methodology can be extended beyond Gaussian states, and generalized to continuous systems and multidimensional arrays with arbitrary networks.

\begin{acknowledgments}
  We acknowledge support from Horizon 2020 QuantERA grant QUOMPLEX, by National Research Council (CNR), Grant 731473, and PRIN PELM (20177PSCKT).
\end{acknowledgments}
\section*{Appendix:Graph and parameters of the model}
\begin{figure*}
  \centering
  \includegraphics[width=0.8\textwidth]{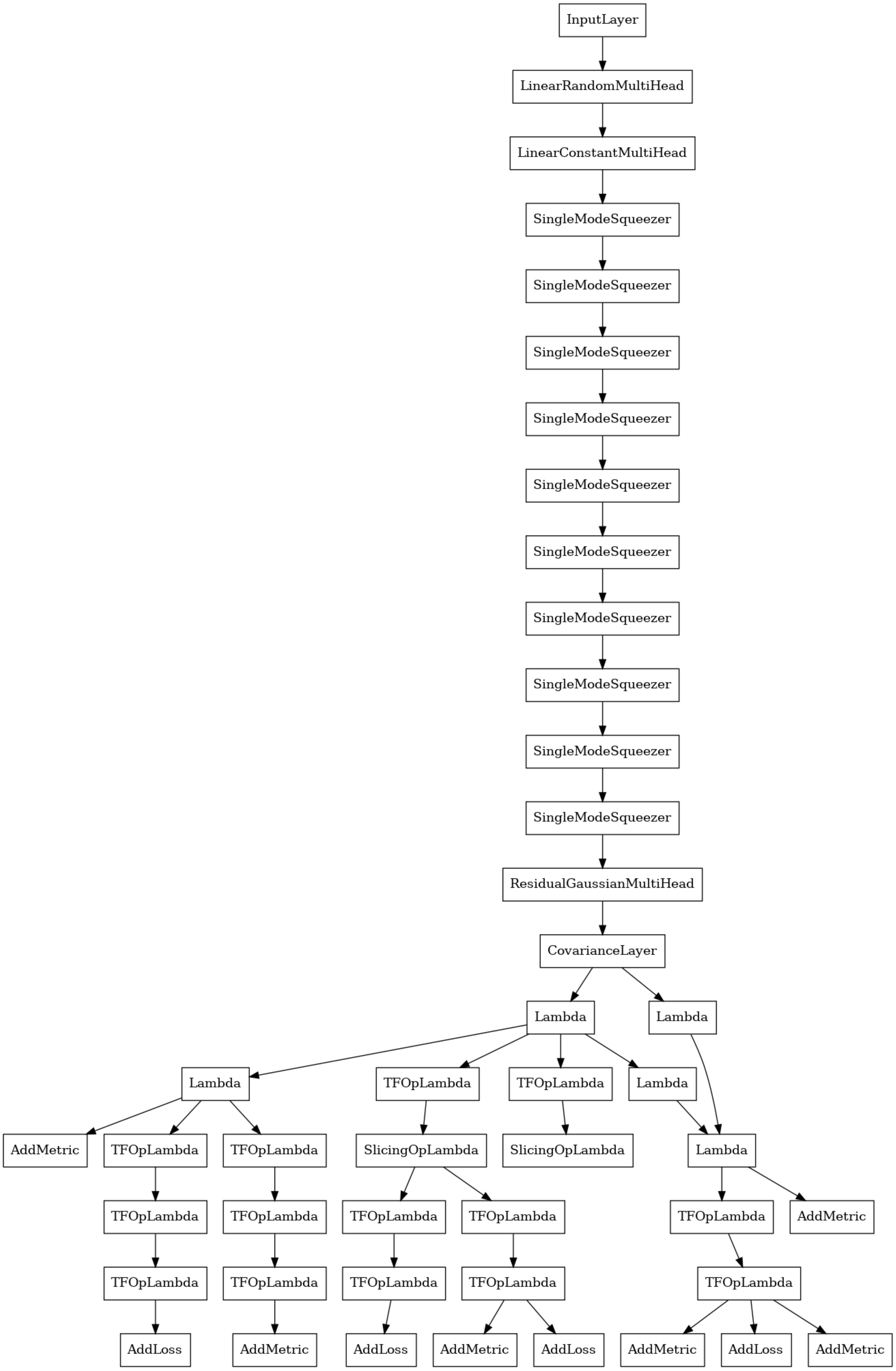}
  \caption{Graph of the multi-head model for the quantum solitons. One can notice the central computational core represented by the cascade of multi-head layers, including the interferometers {\tt LinearRandomMultiHead}, displacements {\tt LinearConstantMultiHead} and squeezing layers {\tt SingleModeSqueezer}. Then we have the activation layer, computing
    the Gaussian characteristic function {\tt ResidualGaussianMultiHead}, and the layer to determine the covariance matrix {\tt CovarianceLayer}. After that a series of ``Lambda'' layers, i.e., simple computing layers
  for the cost functions and the various metrics during the training.\label{fig:model}}
\end{figure*}

Figure~\ref{fig:model} shows a graphical representation of the graph of the model.
We optimize the model in order to minimize the Hamiltonian, the resulting parameters give the circuit that produce a state with the minimum value of energy.
The trainable parameters in the model, i.e., those corresponding to the $n$ displacement layers (each layer with $1$ complex parameter), the $n$ squeezing layers (each layer with $1$ complex parameter), and the unitary matrix representing the trainable interferometer (with $N^2/4$ independent parameters).
Thus the model has $N^{2}/4+N$ independent variables.
Training is done by the Adam algorithm ({\tt TensorFlow v2.7.0}).
Gaussian boson sampling from the trained model is done following~\cite{ContiQMI2021}.
%
%Big picture of the model

%\bibliography{/home/claudio/Dropbox/incorso/bibtex/GIGAbib}% Produces the bibliography via BibTeX.
%apsrev4-2.bst 2019-01-14 (MD) hand-edited version of apsrev4-1.bst
%Control: key (0)
%Control: author (8) initials jnrlst
%Control: editor formatted (1) identically to author
%Control: production of article title (0) allowed
%Control: page (0) single
%Control: year (1) truncated
%Control: production of eprint (0) enabled
%

\end{document}